\documentclass[twocolumn,fp,seceq]{jpsj3}

\usepackage{txfonts}
\usepackage{graphicx}
\usepackage{dcolumn}
\usepackage{bm}
\usepackage[usenames]{color} 
\usepackage{braket}
\usepackage{mathptmx}
\newcommand{\Uell}{u_{\ell {\bm k}}}
\newcommand{\Uellp}{u_{\ell' {\bm k}}}
\newcommand{\Eell}{\varepsilon_{\ell}}
\newcommand{\Eellp}{\varepsilon_{\ell '}}

\title{Orbital Magnetism of Bloch Electrons III. 
Application to Graphene
} 

\author{
Masao \surname{Ogata}
}

\inst{\ 
Department of Physics, University of Tokyo, Hongo, Bunkyo-ku, Tokyo 113-0033, Japan \\
} 

\abst{
The orbital susceptibility for graphene is calculated exactly 
up to the first order with respect to the overlap integrals between neighboring atomic orbitals. 
The general and rigorous theory of orbital susceptibility developed in the preceding paper 
is applied to a model for graphene as a typical two-band model. 
It is found that there are contributions from interband, Fermi surface, and occupied states 
in addition to the Landau--Peierls orbital susceptibility. 
The relative phase between the atomic orbitals on the two sublattices related to the chirality of Dirac cones 
plays an important role. 
It is shown that there are some additional contributions to the orbital susceptibility 
that are not included in the previous calculations using the Peierls phase in the tight-binding model for graphene. 
The physical origin of this difference is clarified in terms of the corrections to the Peierls phase.  
}

\begin{document}
\maketitle

\section{Introduction}

Graphene has a simple and interesting electron system that contains massless chiral 
Dirac electrons (or Weyl electrons) in a two-dimensional honeycomb lattice. 
Its various electronic properties have been explored extensively.\cite{Ando,Neto,Wallace,SlovWeiss} 
Among them, orbital magnetism is an interesting property since 
the Dirac electrons have strong interband effects\cite{Kubo} between the upper and lower 
Dirac cones. 

Actually in early research, McClure\cite{McClure} showed that orbital susceptibility has 
a delta function-like peak as a function of chemical potential $\mu$ where
the two Dirac cones come in contact with each other. 
This is confirmed by using the exact one-line formula (Fukuyama formula).\cite{Fukuyama,FukuGra} 
However, these calculations assume a linear dispersion $\varepsilon=\pm v|{\bm k}|$ 
with a finite cutoff. 
In actual graphene, on the other hand, the energy dispersion deviates from the linear dispersion away 
from the Dirac points in the Brillouin zone. 
Therefore, it is necessary to take account of the Bloch bands. 
Although there have been several studies on the magnetic susceptibility for 
graphene,\cite{Safran,Safran2,Saito,KoshinoAndo,Gomez,Piechon}
we revisit this issue in the present paper by performing a systematic expansion with respect 
to the overlap integrals between nearest-neighbor atomic orbitals based on 
an exact formula expressed in terms of 
Bloch wave functions and the energy dispersion.\cite{OgaFuku1,OgataII} 

The orbital susceptibility for graphene or a two-dimensional honeycomb lattice
was studied\cite{Safran,Safran2,Saito} using the Fukuyama formula\cite{Fukuyama} 
\begin{equation}
\chi = \frac{e^2}{\hbar^2c^2} k_{\rm B} T \sum_{{\bm k},n} {\rm Tr} \ 
\gamma_x {\cal G} \gamma_y {\cal G} \gamma_x {\cal G} \gamma_y {\cal G},
\label{FukuyamaF}
\end{equation}
where $\cal G$ represents the thermal Green's function ${\cal G}({\bm k}, \varepsilon_n)$ 
of the matrix form with respect to the band indices,  $\varepsilon_n$ is the Matsubara frequency, 
and $\gamma_\mu$ ($\mu=x,y$) is the current operator in the $\mu$-direction divided by $e/\hbar$. 
The spin multiplicity of 2 has been taken into account and Tr denotes 
the trace over the band indices. 
This formula is exact if all the bands are taken into account. 
However, in the calculations\cite{Safran,Safran2,Saito},
the band indices of the Green's functions were restricted to the upper and lower Dirac cones. 
Since the band indices of the Green's functions in Eq.~(\ref{FukuyamaF}) should not be 
restricted to a few bands, as discussed in detail in Ref.~15, 
the results should be reexamined. 

Recently, by claiming that there are some \lq\lq correction terms''\cite{KoshinoAndo,Gomez} 
to the exact formula in (\ref{FukuyamaF}) in the case of the two-band model,
the orbital susceptibility for graphene was studied. 
Calculations based on the Peierls phase in the tight-binding model\cite{Piechon} also gave 
the same susceptibility.
More recently, Gao {\it et al}.\cite{Gao} studied a model for gapped graphene based on the 
wave-packet formalism and obtained consistent results with those obtained from the Peierls phase.\cite{Piechon}
Although the above results are consistent with each other, we examine this issue 
motivated by the following findings. 
Recently we studied the orbital susceptibility for single-band models based on an 
exact formalism\cite{OgaFuku1,OgataII} (referred to as I and II in the following).
It was found that there are comparable contributions in addition to the susceptibility originating 
from the Peierls phase.\cite{OgataII}
Furthermore, we clarified the corrections of the Peierls phase in the tight-binding model.\cite{Matsuura} 
Therefore, we expect that there are additional contributions also in graphene. 
For this purpose, 
we think it is important to calculate the susceptibility by systematic expansion with respect to 
the overlap integrals, as carried out for single-band models in II.\cite{OgataII}
 
In our preceding paper I\cite{OgaFuku1},
we rewrote the Fukuyama formula in (\ref{FukuyamaF}) in terms of Bloch wave functions 
and obtained a new and equivalent formula for the orbital susceptibility $\chi$ as follows: 
\begin{equation}
\chi = \chi_{\rm LP} + \chi_{\rm inter} + \chi_{\rm FS} + \chi_{\rm occ}. 
\label{FinalChi}
\end{equation}
The suffixes of $\chi_{\rm LP}$, $\chi_{\rm inter}$, $\chi_{\rm FS}$, and $\chi_{\rm occ}$ denote 
Landau--Peierls, interband, Fermi surface, and occupied states, respectively.\cite{OgaFuku1}
Note that the formula in (\ref{FinalChi}) is exact, as is Eq.~(\ref{FukuyamaF}).
As shown in II\cite{OgataII}, 
$\chi_{\rm LP}$ is in the first order with respect to overlap integrals. 
Thus, we calculate each term in (\ref{FinalChi}) for graphene up to the same order with $\chi_{\rm LP}$. 
In contrast to the previous studies,\cite{Safran,Safran2,Saito,KoshinoAndo,Gomez,Piechon}  
this is the first exact calculation up to the first order with respect to the overlap integrals
starting from the atomic limit. 
We will show that there are some contributions that were not included before. 
The physical origin of these additional contributions is discussed in terms of the corrections to 
the Peierls phase in the tight-binding model.\cite{Matsuura}
In the single-band model studied in II, $\chi_{\rm inter}$ vanishes.\cite{OgataII}
In contrast, in graphene, which is a typical two-band model, we show that $\chi_{\rm inter}$ 
contributes to the total susceptibility. 

This paper is organized as follows. 
In Section 2, we develop an atomic orbital model for graphene. 
We develop a usual tight-binding model but the wave functions are explicitly 
obtained in terms of p$_\pi$ atomic orbitals for carbon atoms. 
Then we calculate the orbital susceptibility in Section 3 
by systematic expansion with respect to the overlap integrals. 
It is found that the relative phase between the atomic orbitals for the A and B sublattices 
plays an important role leading to contributions comparable to $\chi_{\rm LP}$. 
Section 4 is devoted to discussion and summary.

\section{Atomic orbital model for graphene}

\subsection{${\rm p}_\pi$ orbitals and Hamiltonian}

\begin{figure}
\begin{center}
\includegraphics[width=8cm]{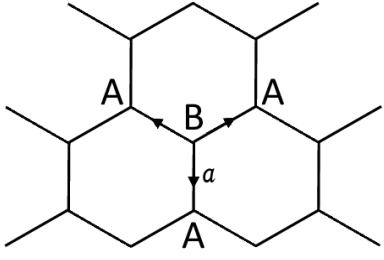}
\vskip -1.5truecm
\caption{Honeycomb lattice for graphene. A and B represent the sublattices and 
the arrows are the vectors from a site on the B sublattice to its nearest-neighbor 
sites on the A sublattice.
$a$ is the distance between the two sites.}
\label{Fig:Gra}
\end{center}
\end{figure}

Since there are two carbon atoms in a unit cell (A and B sublattices) as shown in 
Fig.~\ref{Fig:Gra}, we have two bands that form massless Dirac electrons, or Weyl electrons. 
First, we construct the ${\rm p}_\pi$ (or p$_z$) band for graphene. 
[In the following, we do not consider the contributions from the core-level electrons, 
i.e., the 1s orbital, or the 2s, p$_x$, and p$_y$ orbitals forming $\sigma$-bonds. 
We focus on the contributions from the ${\rm p}_\pi$ orbitals of carbon atoms 
considering that only the ${\rm p}_\pi$ band crosses the Fermi energy.
]
As in I and II,\cite{OgaFuku1,OgataII} we assume that  
the periodic potential $V({\bm r})$ is written as 
\begin{equation}
V({\bm r}) = \sum_{{\bm R}_i} V_0({\bm r}-{\bm R}_i),
\label{PotSum}
\end{equation} 
where ${\bm R}_i$ represents the positions of carbon atoms forming a honeycomb lattice. 
Under the periodic potential $V({\bm r})$, 
wave functions are given by $e^{i{\bm k}\cdot {\bm r}} \Uell({\bm r})$, where 
$\Uell({\bm r})$ satisfies
\begin{equation}
H_{\bm k} \Uell({\bm r}) = \Eell({\bm k}) \Uell({\bm r}),
\label{UellEq}
\end{equation}
with 
\begin{equation}
H_{\bm k} = \frac{\hbar^2 k^2}{2m} - \frac{i\hbar^2}{m} {\bm k}\cdot {\bm \nabla}
- \frac{\hbar^2}{2m} {\bm \nabla}^2 + V({\bm r}).
\label{HamiltonianK}
\end{equation}

For the carbon 2p$_\pi$ orbital, it was discussed that the nucleus charge 
is screened by core electrons, and as a result, the effective 
atomic potential is given by 
\begin{equation}
V_0({\bm r}) = - \frac{Z_{\rm eff} e^2}{r},
\end{equation} 
where $Z_{\rm eff}$ represents the effective charge, $Z_{\rm eff}=3.25$.\cite{Slater}
The ${\rm p}_\pi$ (or ${\rm p}_z$) atomic orbital is then given by
\begin{equation}
\phi_{{\rm p}\pi} ({\bm r}) = \frac{1}{\sqrt{24} (a_{\rm B}^*)^{5/2}} 
\sqrt{\frac{3}{4\pi}} z e^{-r/2a_{\rm B}^*},
\label{PiOrbital}
\end{equation}
where $a^*_{\rm B}$ is the renormalized Bohr radius
\begin{equation}
a^*_{\rm B} = \frac{a_{\rm B}}{Z_{\rm eff}} = \frac{a_{\rm B}}{3.25},
\label{RenormalizedA}
\end{equation}
with $a_{\rm B}=\hbar^2/me^2$. 

As in II,\cite{OgataII} we use the orthogonal wave functions\cite{Lowdwin}
\begin{equation}
\Phi_{{\rm p}\pi} ({\bm r}-{\bm R}_i) = \phi_{{\rm p}\pi} ({\bm r}-{\bm R}_i) - \sum_{j={\rm n.n.}} \frac{s}{2} 
\phi_{{\rm p}\pi} ({\bm r}-{\bm R}_j),
\label{OrthoNormal}
\end{equation}
where the $j$-summation represents the sum over the nearest-neighbor (n.n.) sites 
of ${\bm R}_i$ and $s$ is the overlap integral 
\begin{equation}
s = \int \phi_{{\rm p}\pi}^*({\bm r}-{\bm R}_j) \phi_{{\rm p}\pi}({\bm r}-{\bm R}_i) d{\bm r}. 
\label{defOverlap}
\end{equation}
Note that $s$ is independent of the direction ${\bm R}={\bm R}_j-{\bm R}_i$
since the ${\rm p}_\pi$ orbital is isotropic in the $xy$-plane. 
In the following, we calculate the orbital susceptibility up to the first order with respect 
to \lq\lq overlap integrals" whose integrand contains the overlap of atomic orbitals, 
$\phi_{{\rm p}\pi}^* ({\bm r}-{\bm R}_j) \phi_{{\rm p}\pi} ({\bm r}-{\bm R}_i)$, with ${\bm R}_j \ne {\bm R}_i$. 

Using these orthogonal wave functions, we consider two linear combinations of atomic 
orbitals (LCAOs) for the A and B sublattices defined as
\begin{equation}
\varphi_{{\rm A}{\bm k}}^{\rm ortho} ({\bm r}) = \frac{1}{\sqrt{N}}
\sum_{{\bm R}_{{\rm A}i}} e^{-i{\bm k}({\bm r}-{\bm R}_{{\rm A}i})} 
\Phi_{{\rm p}\pi} ({\bm r}-{\bm R}_{{\rm A}i}),
\end{equation}
and
\begin{equation}
\varphi_{{\rm B}{\bm k}}^{\rm ortho} ({\bm r}) = \frac{1}{\sqrt{N}}
\sum_{{\bm R}_{{\rm B}i}} e^{-i{\bm k}({\bm r}-{\bm R}_{{\rm B}i})} 
\Phi_{{\rm p}\pi} ({\bm r}-{\bm R}_{{\rm B}i}). 
\end{equation}
Here $N$ is the total number of sites on each sublattice and 
${\bm R}_{{\rm A}i}$ (${\bm R}_{{\rm B}i}$) represents the position of the site in  
the A (B) sublattice in the $i$-th unit cell. 

We assume that $\Uell({\bm r})$ can be expanded in terms of 
$\varphi_{{\rm A}{\bm k}}^{\rm ortho}({\bm r})$ and 
$\varphi_{{\rm B}{\bm k}}^{\rm ortho}({\bm r})$. 
The mixing of other orbitals is neglected. In this approximation, 
the matrix elements of $H_{\bm k}$ are given by
\begin{equation}
h_{nm}({\bm k}) = \int \varphi_{n{\bm k}}^{{\rm ortho}*}({\bm r}) H_{\bm k} 
\varphi_{m{\bm k}}^{\rm ortho} ({\bm r}) d{\bm r},
\end{equation}
where $n, m={\rm A}$ or B. 
Considering that the nearest-neighbor sites are A and B sublattices, we obtain\cite{OgataII}
\begin{equation}
\begin{split}
h_{\rm AA}&=h_{\rm BB} = E_{{\rm p}\pi} + C_{{\rm p}\pi {\rm p}\pi}, \cr
h_{\rm AB}&=h_{\rm BA}^* 
= -t\gamma_{\bm k},
\label{HMatrixGr}
\end{split}
\end{equation}
where $E_{{\rm p}\pi}$ is the atomic energy eigenvalue for the p$_\pi$ orbital and 
\begin{equation}
\begin{split}
C_{{\rm p}\pi{\rm p}\pi} &= \int 
\phi_{{\rm p}\pi}^* ({\bm r}) \sum_{{\bm R} \ne 0} V_0({\bm r}-{\bm R}) \phi_{{\rm p}\pi} ({\bm r}) d{\bm r}, \cr
t &= t_0 + s c_{{\rm p}\pi}, \cr
t_0 &=  -\int \phi_{{\rm p}\pi}^* ({\bm r}-{\bm R}) V_0({\bm r}-{\bm R}) \phi_{{\rm p}\pi} ({\bm r}) d{\bm r}, \cr
c_{{\rm p}\pi} &=\int \phi_{{\rm p}\pi}^* ({\bm r}) V_0({\bm r}-{\bm R}) \phi_{{\rm p}\pi} ({\bm r}) d{\bm r}. 
\label{tdef1s}
\end{split}
\end{equation}
The derivations of $t_0$ and $c_{{\rm p}\pi}$ in the hopping integral $t$ 
were discussed and justified in II.\cite{OgataII}
$\gamma_{\bm k}$ is given by 
\begin{equation}
\gamma_{\bm k} = \sum_{{\bm R}} e^{-i{\bm k}\cdot {\bm R}},
\label{DefGamma}
\end{equation}
where ${\bm R}$ are the vectors from a B site to its nearest-neighbor A sites, 
${\bm R}={\bm R}_{{\rm A}j} - {\bm R}_{{\rm B}i}$.   
(Here, we have assumed only the nearest-neighbor hopping integrals, 
but the extension to longer-range hopping integrals is straightforward.) 
Explicitly, $\gamma_{\bm k}$ becomes
\begin{equation}
\gamma_{\bm k} = e^{-ik_y a} + e^{i(\frac{\sqrt{3}}{2} k_x + \frac{1}{2} k_y )a}
+ e^{i(-\frac{\sqrt{3}}{2} k_x + \frac{1}{2} k_y )a}, 
\end{equation}
where $a$ is the distance between the nearest-neighbor sites, i.e., $a=|{\bm R}|$. 

For the ${\rm p}_\pi$-orbital, the integrals can be calculated explicitly as\cite{Mulliken}
\begin{equation}
\begin{split}
s &= \left( 1+{\tilde p} +\frac{2}{5}{\tilde p}^2 +\frac{{\tilde p}^3}{15} \right)\ e^{-{\tilde p}}, \cr
t_0 &= \frac{Z_{\rm eff} e^2}{4a_{\rm B}^*} \left( 1+{\tilde p} +\frac{{\tilde p}^2}{3} \right)\ e^{-{\tilde p}}, \cr
c_{{\rm p}\pi} &= -\frac{Z_{\rm eff} e^2}{4a_{\rm B}^*} 
\left\{ \frac{2}{\tilde p} -\frac{3}{{\tilde p}^3} + \left(1+\frac{4}{\tilde p} 
+\frac{6}{\tilde p^2} +\frac{3}{\tilde p^3} \right) e^{-2\tilde p} \right\}, 
\label{OverlapIntPpi}
\end{split}
\end{equation}
with ${\tilde p}=a/2a_{\rm B}^*$. 
Note that these integrals do not depend on the direction of ${\bm R}$. 
$s$ and $t$ are in the first order with respect to the overlap integrals, which means that they are 
proportional to $e^{-{\tilde p}}$. 
Using $a=1.42$A in graphene, we obtain ${\tilde p}=4.36$, $s=0.237$, and $t=3.55$eV. 
The overlap integrals and various matrix elements are shown in Appendix A. 

\subsection{Energy dispersion and chirality around Dirac points}

By diagonalizing the Hamiltonian in (\ref{HMatrixGr}), 
we obtain normalized eigenfunctions and energy eigenvalues as follows:
\begin{equation}
\begin{split}
u_{{\rm p}\pi {\bm k}}^{+}({\bm r}) &= \frac{i}{\sqrt{2}} \left\{ 
e^{ \frac{i}{2}\theta_{\bm k}} \varphi_{{\rm A}{\bm k}}^{\rm ortho} ({\bm r}) -
e^{-\frac{i}{2}\theta_{\bm k}} \varphi_{{\rm B}{\bm k}}^{\rm ortho} ({\bm r}) \right\}, \cr
&\qquad\qquad {\rm for} \quad \varepsilon_{{\rm p}\pi}^+ ({\bm k}) = E_{{\rm p}\pi} + 
C_{{\rm p}\pi {\rm p}\pi} + \varepsilon_{\bm k}, \cr
u_{{\rm p}\pi {\bm k}}^{-}({\bm r}) &= \frac{1}{\sqrt{2}} \left\{ 
e^{ \frac{i}{2}\theta_{\bm k}} \varphi_{{\rm A}{\bm k}}^{\rm ortho} ({\bm r}) +
e^{-\frac{i}{2}\theta_{\bm k}} \varphi_{{\rm B}{\bm k}}^{\rm ortho} ({\bm r}) \right\}, \cr
&\qquad\qquad {\rm for} \quad \varepsilon_{{\rm p}\pi}^- ({\bm k}) = E_{{\rm p}\pi} + 
C_{{\rm p}\pi {\rm p}\pi} - \varepsilon_{\bm k},
\label{UellGraphene}
\end{split}
\end{equation}
with 
\begin{equation}
\begin{split}
\varepsilon_{\bm k} &= t|\gamma_{\bm k}|, \cr
|\gamma_{\bm k}| &= \sqrt{1+4\cos^2 \frac{\sqrt{3}k_x a}{2}  
+ 4\cos \frac{\sqrt{3}k_x a}{2}   \cos \frac{3k_y a}{2} }, 
\end{split}
\end{equation}
and 
\begin{equation}
e^{i\theta_{\bm k}} = \frac{\gamma_{\bm k}}{|\gamma_{\bm k}|}, 
\qquad (-\pi<\theta_{\bm k}\le \pi). 
\end{equation}
As we can see from Eq.~(\ref{UellGraphene}), $u_{{\rm p}\pi {\bm k}}^{+}({\bm r})$ 
($u_{{\rm p}\pi {\bm k}}^{-}({\bm r})$) is the antibonding (bonding) state between 
the two sublattices with phase factor $e^{\pm i\theta_{\bm k}/2}$.
$\varepsilon_{{\rm p}\pi}^+ ({\bm k})$ ($\varepsilon_{{\rm p}\pi}^- ({\bm k})$) 
gives the energy dispersion of the upper (lower) Dirac cone. 
As in II,\cite{OgataII} the constant energy $E_{{\rm p}\pi} + C_{{\rm p}\pi {\rm p}\pi}$ is 
included in the chemical potential in the following, and we write the Fermi distribution function 
$f(\pm\varepsilon_{\bm k})$ instead of 
$f(E_{{\rm p}\pi} + C_{{\rm p}\pi {\rm p}\pi} \pm\varepsilon_{\bm k})$ for simplicity. 

The phase factors $e^{\pm i\theta_{\bm k}/2}$ in $u_{{\rm p}\pi {\bm k}}^{\pm}({\bm r})$
are determined in order to satisfy 
\begin{equation}
u_{{\rm p}\pi {\bm k}}^{+}(-{\bm r}) = u_{{\rm p}\pi {\bm k}}^{+*}({\bm r}), \quad 
u_{{\rm p}\pi {\bm k}}^{-}(-{\bm r}) = u_{{\rm p}\pi {\bm k}}^{-*}({\bm r}). 
\label{UellSymmetry}
\end{equation}
These relations are required in the case of a centrosymmetric potential and have been used 
in various steps to derive Eq.~(\ref{FinalChi}) in I.\cite{OgaFuku1}
To prove Eq.~(\ref{UellSymmetry}), it is necessary to note that each A site at
${\bm R}_{{\rm A}i}$ has its partner of the B site satisfying $-{\bm R}_{{\rm A}i}={\bm R}_{{\rm B}j}$.
Using this relation, we can see that 
$\varphi_{{\rm A}{\bm k}}^{\rm ortho} (-{\bm r})=\varphi_{{\rm B}{\bm k}}^{{\rm ortho}*} ({\bm r})$ 
holds, which leads to Eq.~(\ref{UellSymmetry}). 

The density of states for the honeycomb lattice was obtained in the context of the phonon
density of states.\cite{DOS}
It is given by 
\begin{equation}
\begin{split}
D(\mu) =& \frac{2\sqrt{|\mu|/t}}{3\sqrt{3} \pi^2 ta^2 } K(\kappa), \qquad {\rm for}\ t< |\mu| <3t, \cr
D(\mu) =& \frac{2\sqrt{|\mu|/t}}{3\sqrt{3} \pi^2 ta^2 \kappa} K(\frac{1}{\kappa}), \qquad {\rm for}\ |\mu|<t,
\label{DOSgra}
\end{split}
\end{equation}
where $K(\kappa)$ is the elliptic integral of the first kind with
\begin{equation}
\kappa = \frac{1}{4} \sqrt{\frac{(1+|\mu|/t)^3 (3-|\mu|/t)}{|\mu|/t}}.
\end{equation}
Note that $E_{{\rm p}\pi} + C_{{\rm p}\pi {\rm p}\pi}$ is included in $\mu$. 
For completeness, its derivation and some limiting cases are given in Appendix B. 

Two Dirac points exist at ${\bm k}={\bm k}_0^\pm =(\pm \frac{4\pi}{3\sqrt{3}a}, 0)$, and 
around these Dirac points, $\gamma_{\bm k}$ can be expanded as
\begin{equation}
\gamma_{\bm k} \sim \mp \frac{3}{2}|{\bm k}-{\bm k}_0^\pm|a\  e^{\pm i \eta},
\label{DiracLimit}
\end{equation}
where $\eta$ is the angle between the $k_x$ axis and vector $\bm k$. 
Thus, the phase $\theta_{\bm k}$ of $\gamma_{\bm k}$ is related to 
the chirality for each Dirac point, and its chirality is opposite for the two Dirac points. 
In the following calculations, we find that the $\bm k$-derivatives of $\theta_{\bm k}$ 
play important roles since $\theta_{\bm k}$ is attached to the wave function
$u_{{\rm p}\pi {\bm k}}^{\pm}({\bm r})$. 
Furthermore, we find that $\partial \theta_{\bm k}/\partial k_\mu$ 
($\mu=x, y$) is related to an integral as 
\begin{equation}
\int u_{{\rm p}\pi {\bm k}}^{\pm \dagger}({\bm r})
\frac{\partial u_{{\rm p}\pi {\bm k}}^{\mp}({\bm r})}{\partial k_\mu} d{\bm r} = \pm \frac{1}{2}\ 
\frac{\partial \theta_{\bm k}}{\partial k_\mu} + O(s^2), 
\label{Berry}
\end{equation}
as shown in Appendix A. 
This integral can be called the 
interband \lq\lq Berry connection'' between the upper and lower Dirac cones. 
However, this kind of integral has been 
familiar for a long time in the literature.\cite{Blount2,WilsonText,HS2}

\section{Orbital susceptibility for Graphene}

Using the energy dispersion in the previous section, the Landau--Peierls susceptibility\cite{Landau,Peierls}
from the ${\rm p}_\pi$ orbital is given by
\begin{equation}
\chi_{\rm LP} = \frac{e^2}{6 \hbar^2 c^2} 
\sum_{{\bm k},\pm} f'(\pm \varepsilon_{\bm k}) 
\left( \varepsilon_{xx} \varepsilon_{yy} - \varepsilon_{xy}^2 \right),
\end{equation}
where we have used the abbreviations 
\begin{equation}
\varepsilon_x = \frac{\partial \varepsilon_{\bm k}}{\partial k_x},  \quad
\varepsilon_{xx} = \frac{\partial^2 \varepsilon_{\bm k}}{\partial k_x^2}, \quad
\varepsilon_{xy} = \frac{\partial^2 \varepsilon_{\bm k}}{\partial k_x \partial k_y}, \ 
{\rm etc.}
\label{AbbrevE}
\end{equation}
Note that 
$\partial \varepsilon^{\pm}_{{\rm p}\pi} ({\bm k})/\partial k_\mu = \pm \partial \varepsilon_{\bm k}/\partial k_\mu
=\pm t\partial |\gamma_{\bm k}|/\partial k_\mu$, 
which is in the first order with respect to the overlap integrals. 
As a result, $\chi_{\rm LP}$ is also in the first order of the overlap integrals. 
In the following, we calculate $\chi_{\rm inter}, \chi_{\rm FS}$, and $\chi_{\rm occ}$ 
up to the same order. 

To evaluate $\chi_{\rm inter}$, $\chi_{\rm FS}$, and $\chi_{\rm occ}$, we use
\begin{equation}
\begin{split}
\frac{\partial u_{{\rm p}\pi {\bm k}}^{\pm}}{\partial k_x} &= \frac{C^\pm}{\sqrt{2N}} 
\biggl[ \sum_{{\bm R}_{{\rm A}i}}  
\left( x-R_{{\rm A}ix} -\frac{1}{2} \theta_x \right) \cr
&\qquad\qquad \times e^{\frac{i}{2}\theta_{\bm k}} e^{-i{\bm k}({\bm r}-{\bm R}_{{\rm A}i})} 
\Phi_{{\rm p}\pi} ({\bm r}-{\bm R}_{{\rm A}i}) \cr
&\quad \mp \sum_{{\bm R}_{{\rm B}i}}  
\left( x-R_{{\rm B}ix} +\frac{1}{2} \theta_x \right) \cr 
&\qquad\qquad \times e^{-\frac{i}{2}\theta_{\bm k}} e^{-i{\bm k}({\bm r}-{\bm R}_{{\rm B}i})} 
\Phi_{{\rm p}\pi} ({\bm r}-{\bm R}_{{\rm B}i}) \biggr], 
\label{Ukderiv}
\end{split}
\end{equation}
with $C^+=1$, $C^- = -i$ 
and we have also used abbreviations such as $\theta_x=\partial \theta_{\bm k}/\partial k_x$. 
In the present model, $\chi_{\rm inter}$ is given by\cite{OgaFuku1,OgataII}
\begin{equation}
\begin{split}
\chi_{\rm inter} &= -\frac{e^2}{\hbar^2 c^2} \sum_{{\bm k}}
\sum_{\varepsilon_\ell=\varepsilon_{{\rm p}\pi}^+ ({\bm k}), \varepsilon_{{\rm p}\pi}^- ({\bm k})} \ 
\sum_{\ell' \ne \ell} \frac{f(\Eell)}{\Eell - \Eellp} \cr
&\times \biggl| \int \frac{\partial \Uell^\dagger}{\partial k_x} 
\left( \frac{\partial H_{\bm k}}{\partial k_y} + \frac{\partial \Eell}{\partial k_y} \right) \Uellp d{\bm r} 
-(x\leftrightarrow y) \biggr|^2, 
\label{ChiInterPpi}
\end{split}
\end{equation}
where the range of the real-space integral $\int \cdots d{\bm r}$ has been extended 
to the whole system size by using the periodicity of $\Uell({\bm r})$,\cite{OgaFuku1}
and $(x\leftrightarrow y)$ 
represents terms in which $x$ and $y$ are exchanged. 
Here, we consider the cases where $\varepsilon_\ell=\varepsilon_{{\rm p}\pi}^+ ({\bm k})$ 
and $\varepsilon_\ell=\varepsilon_{{\rm p}\pi}^- ({\bm k})$. 
Because of the Fermi distribution function $f(\varepsilon_\ell)$, these cases give the 
contributions from the occupied p$_\pi$ orbitals. 
[As discussed before, we do not consider the contributions from the other occupied bands,
i.e., from the 1s, 2s, p$_x$, and p$_y$ orbitals.]
In contrast, we need to take the summation over $\ell'$ ($\ell'\ne\ell$) 
in Eq.~(\ref{ChiInterPpi}),
because these terms originate from the virtual processes in the second-order perturbation. 

In the case of the 1s single-band model discussed in II,\cite{OgataII}
the second line of (\ref{ChiInterPpi}) vanishes both in the zeroth and first order 
with respect to the overlap integrals since $\hat L_z \phi_{\rm 1s}({\bm r})=0$. 
However, in the present two-band model, 
there is a contribution in the {\it zeroth order} that involves the derivatives of the phase 
$\theta_{\bm k}$ {\it even if} $\hat L_z \phi_{{\rm p}_\pi}({\bm r})=0$. 
This is in sharp contrast to the single-band case. 
Since there is a zeroth-order term, the infinite summation of $\ell'$ in $\chi_{\rm inter}$ should be carried out carefully
to obtain the contributions up to the first order of the overlap integrals. 

Furthermore, it should be noted that the denominator in (\ref{ChiInterPpi}) becomes 
$\varepsilon_{{\rm p}\pi}^\pm ({\bm k})-\varepsilon_{{\rm p}\pi}^\mp ({\bm k}) = 
\pm 2\varepsilon_{\bm k}$ when $\varepsilon_{\ell}=\varepsilon_{{\rm p}\pi}^\pm ({\bm k})$ 
and $\varepsilon_{\ell'}=\varepsilon_{{\rm p}\pi}^\mp ({\bm k})$.
Since this denominator is in the first order with respect to the overlap integrals,  
this case should be treated carefully. 
As shown in Appendix C, however, the numerator in this case 
turns out to be proportional to the fourth order with respect to the overlap integrals. 
Therefore, the perturbation does not break down.  
In this case, we find that the combination of 
$\frac{\partial H_{\bm k}}{\partial k_y} + \frac{\partial \Eell}{\partial k_y}$ in 
Eq.~(\ref{ChiInterPpi}) is important. 

As shown in Appendix C, 
the summation over $\ell'$ 
in (\ref{ChiInterPpi}) is carried out analytically, and we obtain 
\begin{equation}
\begin{split}
&\chi_{\rm inter} = \frac{e^2}{\hbar^2 c^2} \sum_{{\bm k}, \pm} f(\pm \varepsilon_{\bm k})
\biggl[ \frac{\hbar^2}{8m} (1\pm 2s|\gamma_{\bm k}| ) (\theta_x^2+\theta_y^2) \cr
&\pm 3a_{\rm B}^{*2} \varepsilon_{\bm k} (\theta_x^2 + \theta_y^2 )
\pm \frac{1}{8} (\varepsilon_{xx} \theta_y^2 - 2\varepsilon_{xy} \theta_x \theta_y 
+ \varepsilon_{yy} \theta_x^2 ) 
\biggr] + O(s^2). 
\label{GrChiInterFin}
\end{split}
\end{equation}
Note that all the terms are related to the derivatives of $\theta_{\bm k}$. 
In the single-band case,\cite{OgataII} the relative phase $\theta_{\bm k}$ does 
not appear and thus $\chi_{\rm inter}$ vanishes. 

Next we calculate $\chi_{\rm FS}$ from the p$_\pi$ orbital,  
which is given by\cite{OgaFuku1,OgataII}
\begin{equation}
\begin{split}
\chi_{\rm FS} &= \frac{e^2}{\hbar^2 c^2} \sum_{{\bm k},\pm} f'(\pm \varepsilon_{\bm k}) \biggl\{ 
\pm \varepsilon_x
\int \frac{\partial u_{{\rm p}\pi {\bm k}}^{\pm \dagger}}{\partial k_y} 
\left( \frac{\partial H_{\bm k}}{\partial k_x} \pm \varepsilon_x  \right)
\frac{\partial u_{{\rm p}\pi {\bm k}}^{\pm}}{\partial k_y}  d{\bm r} \cr
&\qquad \mp \varepsilon_x 
\int \frac{\partial u_{{\rm p}\pi {\bm k}}^{\pm \dagger}}{\partial k_x} 
\left( \frac{\partial H_{\bm k}}{\partial k_y} \pm \varepsilon_y \right)
\frac{\partial u_{{\rm p}\pi {\bm k}}^{\pm}}{\partial k_y}  d{\bm r} \biggr\} + (x\leftrightarrow y). 
\label{ChiFSPpi}
\end{split}
\end{equation}
These integrals can be rewritten with the help of the relation $\hat L_z \phi_{{\rm p}_\pi}({\bm r})=0$ 
in a similar way to $\chi_{\rm inter}$. Then we obtain
\begin{equation}
\begin{split}
\chi_{\rm FS} &= \frac{e^2}{\hbar^2 c^2} \sum_{{\bm k}, \pm} f'(\pm \varepsilon_{\bm k}) 
\biggl[  \left( 6a_{\rm B}^{*2} + \frac{\hbar^2 s}{4mt}\right) (\varepsilon_x^2 + \varepsilon_y^2 )  \cr
&+ \frac{1}{4} \left( \varepsilon_x^2 \theta_y^2 
-2\varepsilon_x \varepsilon_y \theta_x \theta_y + \varepsilon_y^2 \theta_x^2 \right) \cr
&-\frac{\varepsilon_{\bm k}}{4} (\varepsilon_x \theta_x \theta_{yy} -\varepsilon_x \theta_y \theta_{xy} 
-\varepsilon_y \theta_x \theta_{xy} + \varepsilon_y \theta_y \theta_{xx}) \biggr] + O(s^2). 
\label{GrChiFSFin}
\end{split}
\end{equation}
Details of the derivation are shown in Appendix C. 

Finally, $\chi_{\rm occ}$ is given by
\begin{equation}
\begin{split}
\chi_{\rm occ} &= -\frac{e^2}{2\hbar^2 c^2}
\sum_{{\bm k}, \pm} f(\pm \varepsilon_{\bm k}) \biggl\{ 
\pm \varepsilon_{xy} 
\int \frac{\partial u_{{\rm p}\pi {\bm k}}^{\pm \dagger}}{\partial k_x} 
\frac{\partial u_{{\rm p}\pi {\bm k}}^{\pm}}{\partial k_y} d{\bm r} \cr
&\qquad +\left( \frac{\hbar^2}{m} \mp \varepsilon_{xx} \right)
\int \frac{\partial u_{{\rm p}\pi {\bm k}}^{\pm \dagger}}{\partial k_y} 
\frac{\partial u_{{\rm p}\pi {\bm k}}^{\pm}}{\partial k_y} d{\bm r}
\biggr\} + (x\leftrightarrow y).
\label{ChiC}
\end{split}
\end{equation}
[Here again, we do not consider the contributions from the other occupied bands.]
In a similar way carried out in our preceding paper,\cite{OgataII}
$\chi_{\rm occ}$ becomes  
\begin{equation}
\begin{split}
\chi_{\rm occ} &= -\frac{e^2}{2\hbar^2 c^2} \sum_{{\bm k}, \pm} f(\pm \varepsilon_{\bm k}) 
\biggl[ \pm \frac{1}{4} \varepsilon_{xy} \theta_x \theta_y \cr
& + \left( \frac{\hbar^2}{m} \mp \varepsilon_{xx} \right)
\left( \langle y^2 \rangle_{{\rm p}\pi {\rm p}\pi} + \frac{1}{4} \theta_y^2 \right) \cr
&\mp \frac{\hbar^2}{m} {\rm Re} \sum_{{\bm R}} e^{-i\theta_{\bm k}} e^{-i{\bm k}\cdot {\bm R}}
\langle y^2 \rangle_{R, {\rm p}\pi {\rm p}\pi} \biggr] +O(s^2) + (x\leftrightarrow y), 
\label{GrChiOcc}
\end{split}
\end{equation}
where Re denotes the real part and the expectation value is defined as 
\begin{equation}
\langle {\cal O} \rangle_{R,{\rm p}\pi {\rm p}\pi}
= \int \Phi_{{\rm p}\pi}^* ({\bm r}-{\bm R}) {\cal O} \Phi_{{\rm p}\pi} ({\bm r}) d{\bm r}.
\label{Odefinition}
\end{equation}
Note that there is a difference between $\Phi_{{\rm p}\pi}({\bm r})$ 
and $\phi_{{\rm p}\pi}({\bm r})$ as in Eq.~(\ref{OrthoNormal}). 
Using the expectation values in (\ref{x2in2ppi}), 
we obtain
\begin{equation}
\begin{split}
\chi_{\rm occ} &= \frac{e^2}{\hbar^2 c^2} \sum_{{\bm k}, \pm} f(\pm \varepsilon_{\bm k}) 
\biggl[ -\frac{\hbar^2}{m} \left\{ 6 a_{\rm B}^{*2} + \frac{1}{8} (\theta_x^2 +\theta_y^2) \right\} \cr
&\pm 3a_{\rm B}^{*2} (\varepsilon_{xx} + \varepsilon_{yy}) 
\pm \frac{1}{8} (\varepsilon_{xx} \theta_y^2 - 2\varepsilon_{xy} \theta_x \theta_y 
+ \varepsilon_{yy} \theta_x^2 ) \cr
&\pm \frac{\hbar^2 \varepsilon_{\bm k}}{2mt} 
\langle x^2 + y^2 \rangle_{R, {\rm p}\pi {\rm p}\pi} \biggr] + O(s^2), 
\label{GrChiOccFin}
\end{split}
\end{equation}
where we have used the definition of $\gamma_{\bm k}$ in (\ref{DefGamma}). 

\begin{figure}
\begin{center}
\includegraphics[width=10cm]{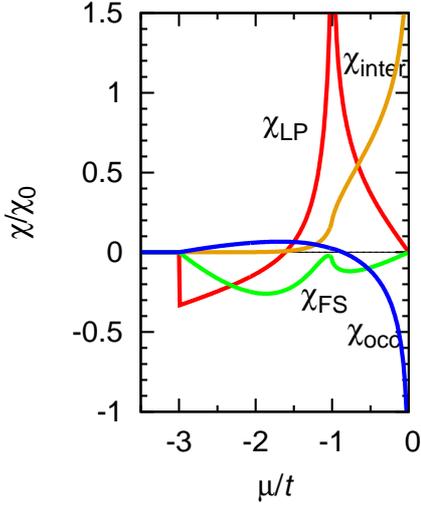}
\caption{
(Color online) Each contribution of orbital susceptibility, 
$\chi_{\rm LP}$, $\chi_{\rm inter}$, $\chi_{\rm FS}$, 
and $\chi_{\rm occ}$ as a function of chemical potential $\mu$ at $T=0$ 
in the case of graphene or a two-dimensional honeycomb lattice.
Each contribution is normalized by the Pauli susceptibility $\chi_0$ at the band edge
(see the text).}
\label{Fig:GraChiAll}
\end{center}
\end{figure}

We numerically calculate $\chi_{\rm LP}$, $\chi_{\rm inter}$, $\chi_{\rm FS}$, 
and $\chi_{\rm occ}$ as a function of chemical potential $\mu$ at $T=0$. 
The results are shown in Fig.~\ref{Fig:GraChiAll}, in which each contribution is 
normalized by the Pauli susceptibility $\chi_0$ at the band edge ($\mu=\pm 3t$), i.e., 
\begin{equation}
\chi_0 = \frac{3e^2}{8\pi \hbar^2 c^2} ta^2 L^2,
\end{equation}
where the system size is $L^2=3\sqrt{3}a^2 N/2$ with $N$ being the total number of unit cells. 
Here, we have used the fact that the model is equivalent to free electrons with effective 
mass $m^*=2\hbar^2/3ta^2$ at the bottom of the band. 

There are several remarks on the above results. 

(1) In contrast to the single-band case discussed in I\cite{OgaFuku1}, there appear several terms 
involving $\theta_x$ and $\theta_y$, which originate from the two-band nature.  
Furthermore, $\chi_{\rm inter}$ is non-zero, which is in sharp contrast to the single-band case, 
where $\chi_{\rm inter}$ vanishes. 
This is also owing to the two-band nature. 

(2) There are three terms that are in the zeroth order with respect to the 
overlap integrals: the first term of $\chi_{\rm inter}$ in Eq.~(\ref{GrChiInterFin}) 
and the first two terms of $\chi_{\rm occ}$ in Eq.~(\ref{GrChiOccFin}). 
However, the first term of $\chi_{\rm inter}$ and the second term of $\chi_{\rm occ}$, 
both of which diverge as $\mu\rightarrow 0$, exactly cancel with each other. 
Therefore, only the first term of $\chi_{\rm occ}$ contributes to the total susceptibility 
in the zeroth order. 
This term is proportional to the electron number in the p$_\pi$ band, i.e., 
\begin{equation}
\chi_{\rm occ:1} \equiv - \frac{3e^2a_{\rm B}^{*2}}{m c^2} n(\mu),
\label{IntraDia}
\end{equation}
where $n(\mu)$ represents the total electron number with spin degeneracy 
when the chemical potential is $\mu$, and it is calculated from the density of states 
in (\ref{DOSgra}) as $n(\mu)/L^2= 2\int_{-3t}^\mu D(\mu) d\mu$. 
$\chi_{\rm occ:1}$ represents the contributions from the occupied states in the 
partially filled p$_\pi$-band, which we call {\it intraband atomic diamagnetism}.\cite{OgataII}
Since $\chi_{\rm occ:1}$ does not have a factor of $e^{-{\tilde p}}$, it gives comparative 
contributions as $\chi_{\rm LP}$ in a similar way to the single-band case 
studied in II.

(3) At the band bottom ($\mu=-3t$), only $\chi_{\rm LP}$ has a contribution. 
Its value is just equal to $-1/3\chi_0$, 
which is understood as Landau's diamagnetic orbital susceptibility for free electrons. 
Furthermore, $\chi_{\rm LP}$  
has a diverging peak at $\mu=-t$, which corresponds to the van Hove singularity. 

(4) $\chi_{\rm FS}$ is always negative and its small wiggle at $\mu=-t$ is owing to 
a subtle cancellation between the last two terms in (\ref{GrChiFSFin}), 
both of which diverge as $\mu\rightarrow 0$. 
$\chi_{\rm FS}$ has a sizable contribution in the region of $-3t<\mu<-t$. 

\begin{figure}
\hspace{0.2truecm}
\includegraphics[width=9.8cm]{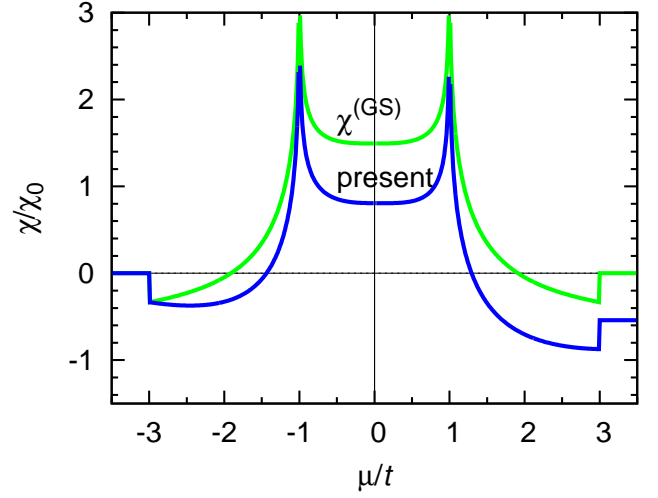}
\caption{
(Color online) Orbital susceptibility as a function of chemical potential $\mu$ 
at $T=0$ in the case of graphene or a two-dimensional honeycomb lattice, 
normalized by the Pauli susceptibility $\chi_0$ at the band edge.
For comparison, $\chi^{({\rm GS})}$ obtained previously\cite{Gomez} is also shown.
A $\delta$-function-like peak at $\mu=0$\cite{McClure} is not included.}
\label{Fig:GraChiComp}
\end{figure}

The total susceptibility $\chi=\chi_{\rm LP}+\chi_{\rm inter}+\chi_{\rm FS}+\chi_{\rm occ}$ 
is shown in Fig.~\ref{Fig:GraChiComp} as a function of $\mu$. 
A $\delta$-function-like peak at $\mu=0$\cite{McClure} is not included, which we discuss shortly. 
It is rather surprising that the total of each contribution
becomes a smooth function of $\mu$ irrespective of the irregular $\mu$-dependences of 
$\chi_{\rm inter}$ and $\chi_{\rm FS}$ near $\mu=-t$ in Fig.~\ref{Fig:GraChiAll}. 
In Fig.~\ref{Fig:GraChiComp}, the present result is compared with 
that obtained by G\`omez-Santos and Stauber,\cite{Gomez} denoted as $\chi^{({\rm GS})}$. 
Apparently, there is a sizable difference from $\chi^{({\rm GS})}$, 
which is discussed in the next section. 

In the present results, there is asymmetry with respect to the sign 
change of $\mu$.  
This is because the contribution $\chi_{{\rm occ}:1}$ in Eq.~(\ref{IntraDia}) 
is a monotonically decreasing function of $\mu$. 
When the p$_\pi$ band is fully filled (i.e., $\mu>3t$), only $\chi_{{\rm occ}:1}$ gives 
the contribution to the orbital susceptibility. 
Therefore, the total susceptibility becomes
\begin{equation}
\chi (\mu>3t) = - \frac{3e^2a_{\rm B}^{*2}}{m c^2} N_e,
\label{AtomicDiaforPpi}
\end{equation}
with $N_e$ being the total electron number of the p$_\pi$ band. 
This is nothing but the atomic diamagnetism from the p$_\pi$ electrons. 

\section{Discussion and Summary}

First, in order to see the relationship between the present result and the previous ones, 
we rewrite the total susceptibility in a different way. 
Using integration by parts for $\chi_{\rm FS}$ in (\ref{GrChiFSFin}), 
we find that the total susceptibility can be rewritten as the 
following simple form:
\begin{equation}
\chi = \chi_{\rm occ:1} + \chi_{\rm LP} + \chi_1 + \chi_2,
\label{ChiAllFinal}
\end{equation}
with 
\begin{equation}
\begin{split}
\chi_1 =& \frac{e^2}{2\hbar^2 c^2} \sum_{{\bm k}, \pm} f(\pm \varepsilon_{\bm k}) 
\biggl[ \pm \varepsilon_{\bm k} \left( \theta_{xx} \theta_{yy} - \theta_{xy}^2 \right) \cr
&\pm (\varepsilon_{x} \theta_x \theta_{yy} 
+ \varepsilon_{y} \theta_y \theta_{xx} - \varepsilon_{x} \theta_y \theta_{xy} 
- \varepsilon_{y} \theta_x \theta_{xy}) \biggr], 
\label{Chi1Final}
\end{split}
\end{equation}
\begin{equation}
\begin{split}
\chi_2=&\frac{e^2}{\hbar^2 c^2} \sum_{{\bm k}, \pm} f(\pm \varepsilon_{\bm k})
\biggl[ \pm ( \varepsilon_{\bm k} \theta_x^2+ \varepsilon_{\bm k} \theta_y^2
-\varepsilon_{xx}- \varepsilon_{yy} ) \cr
&\qquad \times \left( 3a_{\rm B}^{*2} + \frac{\hbar^2 s}{4mt} \right) 
\pm \frac{\hbar^2 \varepsilon_{\bm k}}{2mt} 
\langle x^2 + y^2 \rangle_{R, {\rm p}\pi {\rm p}\pi} \biggr]. 
 \end{split}
\end{equation}
Note that only $\chi_{\rm occ:1}$ is in the zeroth order with respect to the overlap integrals, 
while the other three contributions are in the first order. 
$\chi_{\rm occ:1}$ originates from $\chi_{\rm occ}$, but 
$\chi_1$ and $\chi_2$ are combinations of $\chi_{\rm inter}$, $\chi_{\rm FS}$, 
and $\chi_{\rm occ}$. 
Finally, when we use the interesting relation
\begin{equation}
\varepsilon_{\bm k} \theta_x^2+ \varepsilon_{\bm k} \theta_y^2
-\varepsilon_{xx}- \varepsilon_{yy}  = a^2 \varepsilon_{\bm k},
\label{XRelation}
\end{equation}
(see Appendix D), $\chi_2$ can be rewritten as
\begin{equation}
\chi_2= \frac{e^2}{\hbar^2 c^2} \sum_{{\bm k}, \pm} 
(\pm b \varepsilon_{\bm k})f(\pm \varepsilon_{\bm k}), 
\label{Chi2Final}
\end{equation}
with 
\begin{equation}
\begin{split}
b &= 3a^2 a_{\rm B}^{*2} + \frac{\hbar^2 s}{4mt}a^2 + \frac{\hbar^2}{2mt} 
\langle x^2 + y^2 \rangle_{R, {\rm p}\pi {\rm p}\pi} \cr
&= 3a^2 a_{\rm B}^{*2} + \frac{\hbar^2}{2mt} 
\left( \langle x^2 + y^2 \rangle_{R, {\rm p}\pi {\rm p}\pi}^{(0)} 
- s \langle x^2 + y^2 \rangle_{{\rm p}\pi {\rm p}\pi}^{(0)} \right),
\label{Chi2FinalB}
\end{split}
\end{equation}
where we have used the relation in (\ref{ExpR2inR}). 
Expectation values for an operator $\cal O$
in terms of $\phi_{{\rm p}\pi}({\bm r})$ are defined as 
\begin{equation}
\begin{split}
\langle {\cal O} \rangle_{{\rm p}\pi {\rm p}\pi}^{(0)} 
&= \int \phi_{{\rm p}\pi}^* ({\bm r}) {\cal O} \phi_{{\rm p}\pi} ({\bm r}) d{\bm r}, \cr
\langle {\cal O} \rangle_{R, {\rm p}\pi {\rm p}\pi}^{(0)} 
&= \int \phi_{{\rm p}\pi}^* ({\bm r}-{\bm R}) {\cal O} \phi_{{\rm p}\pi} ({\bm r}) d{\bm r}. 
\label{ExpValueZero}
\end{split}
\end{equation}
Using the expectation values calculated in Appendix A, each term in Eq.~(\ref{Chi2FinalB}) becomes
\begin{equation}
\begin{split}
&3a^2 a_{\rm B}^{*2} = 0.0395 a^4, \cr
&\frac{\hbar^2}{2mt} \langle x^2 + y^2 \rangle_{R, {\rm p}\pi {\rm p}\pi}^{(0)} = 0.0624 a^4, \cr
-&\frac{\hbar^2}{2mt} s \langle x^2 + y^2 \rangle_{{\rm p}\pi {\rm p}\pi}^{(0)} 
= -\frac{6\hbar^2}{mt} s a_{\rm B}^{*2} = -0.0198 a^4,
\end{split}
\end{equation}
when ${\tilde p}=a/2a_{\rm B}^*=4.36$. 

\begin{figure}
\hspace{0.2truecm}
\includegraphics[width=9.8cm]{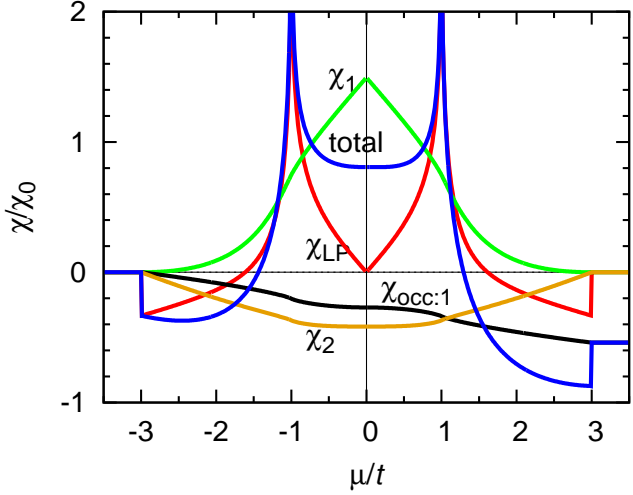}
\caption{
(Color online) Contributions of $\chi_{\rm occ:1}$, $\chi_{\rm LP}$, $\chi_1$, and $\chi_2$ 
together with the total susceptibility as a function of chemical potential $\mu$ 
at $T=0$ in the case of graphene
normalized by the Pauli susceptibility $\chi_0$ at the band edge.}
\label{Fig:GraChiDevide}
\end{figure}

Each contribution, $\chi_{\rm occ:1}$, $\chi_{\rm LP}$, $\chi_1$, and $\chi_2$, 
is plotted in Fig.~\ref{Fig:GraChiDevide} together with the total susceptibility. 
The flat part of the total susceptibility near $\mu=0$ originates from the cancellation 
between the $\mu$-dependences of $\chi_{\rm LP}$ and $\chi_1$. 
As discussed before, only $\chi_{\rm occ:1}$ has asymmetry with respect to $\pm \mu$. 

Here we comment on the $\delta$-function-like peak at $\mu=0$,\cite{McClure} which is not included in 
Figs.~\ref{Fig:GraChiComp} and \ref{Fig:GraChiDevide}. 
As shown by Fukuyama\cite{FukuGra}, 
if the energy dispersion is completely linear around a Dirac point, i.e., $\varepsilon=\pm v|{\bm k}|$, 
we expect 
\begin{equation}
\chi^{({\rm Dirac})} = - \frac{2e^2 v^2}{3\pi^2 \hbar^2 c^2} \frac{\Gamma}{\mu^2+\Gamma^2}, 
\label{DiracMcClure}
\end{equation}
as a contribution from the Dirac cones at $T=0$. 
Here $\Gamma$ is phenomenologically introduced damping and the presence of 
two Dirac points has been taken into account. 
This result at $T=0$ with finite damping\cite{FukuGra} corresponds 
to the $\delta$-function-like peak obtained by McClure\cite{McClure} in clean systems at finite temperatures. 
This peak will originate from the singular behavior of the $\bm k$-derivatives of $\theta_{\bm k}$
at the Dirac points. 
In the present model, we have $v=3ta/2$ from Eq.~(\ref{DiracLimit}) and thus 
$\chi^{({\rm Dirac})}$ is proportional to $t^2$, 
which means that $\chi^{({\rm Dirac})}$ will appear in the second order with respect to the overlap integrals. 
Although $\chi^{({\rm Dirac})}$ is in the second order, it should be included in the total susceptibility 
because of its singular behavior. 

Let us compare the present result with previous ones. G\`omez-Santos and Stauber\cite{Gomez} used
the following formula for orbital susceptibility: 
\begin{equation}
\begin{split}
&\chi^{({\rm GS})} = \frac{e^2}{\hbar^2c^2} k_{\rm B} T \sum_{{\bm k},n} {\rm Tr} \biggl[ 
{\hat \gamma}_x {\cal G} {\hat \gamma}_y {\cal G} {\hat \gamma}_x {\cal G} {\hat \gamma}_y {\cal G} \cr
&\qquad \qquad \qquad 
+\frac{1}{2} \left( {\hat \gamma}_x {\cal G} {\hat \gamma}_y {\cal G} + 
 {\hat \gamma}_y {\cal G} {\hat \gamma}_x {\cal G} \right) {\hat \gamma}_{xy} {\cal G} \biggr],
\label{GomezF}
\end{split}
\end{equation}
where $\cal G$ is now a $2\times 2$ matrix, ${\cal G}=(i\varepsilon_n - {\cal H}_{\bm k})^{-1}$, and 
${\hat \gamma}_\mu = \partial  {\cal H}_{\bm k}/\partial k_\mu$, etc., with 
${\cal H}_{\bm k}$ being the Hamiltonian in a $2\times 2$ matrix form. 
(The spin degeneracy has been included.)
The second term in (\ref{GomezF}) is the \lq\lq correction term"\cite{Gomez} 
added to the exact formula of (\ref{FukuyamaF}).

Actually, before G\`omez-Santos and Stauber, Koshino and Ando\cite{KoshinoAndo} used 
another formula,
\begin{equation}
\begin{split}
&\chi^{({\rm KA})} = \frac{e^2}{2\hbar^2c^2} k_{\rm B} T \sum_{{\bm k},n} {\rm Tr} \biggl[ 
{\hat \gamma}_x {\cal G} {\hat \gamma}_y {\cal G} {\hat \gamma}_x {\cal G} {\hat \gamma}_y {\cal G} \cr
&\quad -2{\hat \gamma}_x {\cal G} {\hat \gamma}_x {\cal G} {\hat \gamma}_y {\cal G} {\hat \gamma}_y {\cal G} 
-\frac{1}{2} {\hat \gamma}_y {\cal G} {\hat \gamma}_{xx} {\cal G} {\hat \gamma}_y {\cal G}
-\frac{1}{2} {\hat \gamma}_x {\cal G} {\hat \gamma}_{yy} {\cal G} {\hat \gamma}_x {\cal G} \biggr],
\label{KoshinoAndoF}
\end{split}
\end{equation}
although they did not calculate $\chi^{({\rm KA})}$ explicitly. 
The last three terms are their \lq\lq correction terms".
We can show that $\chi^{({\rm GS})}$ and $\chi^{({\rm KA})}$ 
are equivalent using the relation 
$\partial  {\cal G}_{\bm k}/\partial k_\mu={\cal G}{\hat \gamma}_\mu {\cal G}$ 
and integration by parts. 

In the present notation, the $2\times 2$ Hamiltonian is given by
\begin{equation}
{\cal H}_{\bm k} = \left( 
\begin{matrix}
0 & -t \gamma_{\bm k} \cr
 -t \gamma_{\bm k}^*  & 0 
\end{matrix}
\right) =  -t \left( {\rm Re} \gamma_{\bm k} \sigma_x - {\rm Im} \gamma_{\bm k} \sigma_y \right), 
\end{equation}
with $\sigma_{x,y}$ being the Pauli matrix. 
After some algebra, we can show that $\chi^{({\rm GS})}$ is given in the present notation as
\begin{equation}
\begin{split}
\chi^{({\rm GS})} &= \frac{e^2}{6\hbar^2 c^2} \sum_{{\bm k}, \pm} 
f'(\pm \varepsilon_{\bm k}) (\varepsilon_{xx} \varepsilon_{yy} - \varepsilon_{xy}^2) \cr
&-\frac{e^2}{4\hbar^2 c^2} \sum_{{\bm k}, \pm} f'(\pm \varepsilon_{\bm k})
\biggl\{ 2\varepsilon_{x} \varepsilon_{y} \theta_x \theta_y  \cr
&\qquad\quad + \varepsilon_{\bm k} (\varepsilon_{x} \theta_x \theta_{yy} 
+ \varepsilon_{y} \theta_y \theta_{xx} -\varepsilon_{x} \theta_y \theta_{xy} 
-\varepsilon_{y} \theta_x \theta_{xy}) \biggr\} \cr
&-\frac{e^2}{2\hbar^2 c^2} \sum_{{\bm k}, \pm} 
\pm f(\pm \varepsilon_{\bm k}) 
\left( \varepsilon_{x} \theta_y \theta_{xy} +\varepsilon_{y} \theta_x \theta_{xy} 
+ \varepsilon_{xy} \theta_x \theta_y \right) .
\label{Gomez}
\end{split}
\end{equation}
When we perform integration by parts in the second summation,
we can see that 
$\chi^{({\rm GS})}$ is exactly equal to $\chi_{\rm LP}+\chi_1$ in Eq.~(\ref{ChiAllFinal}).
Therefore, the difference between the present result and $\chi^{({\rm GS})}$ 
is simply $\chi_{{\rm occ}:1}$ and $\chi_2$. 

The origin of this difference will be the following. In $\chi^{({\rm GS})}$,
(i) the effect of deformation of the wave function $\Uell$ is not taken into account completely 
and (ii) the occupied electron contributions in $\chi_{\rm occ}$ are not included.
(iii) As shown in I,\cite{OgaFuku1} the important interband contributions are taken into 
account through the $f$-sum rule. 
However, \lq\lq the correction terms" introduced in the previous studies 
do not sufficiently take account of these contributions. 

In $\chi^{({\rm GS})}$ and $\chi^{({\rm KA})}$, the effect of a magnetic field 
was introduced in the energy dispersion of the Bloch bands. 
The correct procedure should be to introduce the effect of a magnetic field into the 
Bloch equation in (\ref{UellEq}). 
As a result, the exact Fukuyama formula in Eq.~(\ref{FukuyamaF}) is obtained, 
which contains all the contributions including the effects of deformation of the wave functions.  

As explained in the introduction, calculations based on the Peierls phase\cite{Piechon} in the tight-binding model
give the same susceptibility as $\chi^{({\rm GS})}$.
The equivalence of these two methods is shown in Appendix E. 
This result means that the calculations based on the Peierls phase do not contain 
$\chi_{{\rm occ}:1}$ and $\chi_2$. 
Therefore, we expect that the effects of a magnetic field other than the Peierls phase will 
play important roles. 
Recently, we clarified corrections to the Peierls phase argument in the 
tight-binding model and studied the cases of a benzene molecule and a square lattice.\cite{Matsuura}  
A similar argument can be applied to graphene. 
By extending Pople's formulation on the effects of a magnetic field to the tight-binding model,\cite{Pople}  
we showed that the hopping integral $t$ has the correction 
$\Delta t= t_2 \tilde h^2$ with $\tilde h = eHa^2/2c\hbar$.\cite{Matsuura} 
(The expression for $t_2$ is shown shortly.) 
This correction originates from the deformation of the wave function by the magnetic field.
Since a change in the hopping integral causes a change in the kinetic energy, it contributes to 
the susceptibility as
\begin{equation}
\Delta \chi = \frac{e^2}{\hbar^2 c^2} \sum_{{\bm k}, \pm} 
\left(\pm \frac{a^4 t_2}{t} \varepsilon_{\bm k} \right) f(\pm \varepsilon_{\bm k}). 
\label{Matsuura}
\end{equation}
This formula is equivalent to $\chi_2$ in Eq.~(\ref{Chi2Final}), and 
$b$ in (\ref{Chi2Final}) corresponds to $a^4t_2/t$. 
In the present notation, $t_2$ is given by\cite{Matsuura}
\begin{equation}
\begin{split}
t_2 &= \frac{\hbar^2}{2ma^4} 
\left( \langle x^2 + y^2 \rangle_{R, {\rm p}\pi {\rm p}\pi}^{(0)} 
- s \langle x^2 + y^2 \rangle_{{\rm p}\pi {\rm p}\pi}^{(0)} \right) \cr
&+\frac{1}{2a^4} \biggl[ \langle (R_x y- R_y x)^2 \rangle_{R, {\rm p}\pi {\rm p}\pi}^{(0)}
\langle V_0({\bm r}-{\bm R}) \rangle_{{\rm p}\pi {\rm p}\pi}^{(0)} \cr
&\qquad \qquad -\langle (R_x y- R_y x)^2 V_0({\bm r}-{\bm R})
\rangle_{R, {\rm p}\pi {\rm p}\pi}^{(0)} \biggr].
\label{MatsuuraT2}
\end{split}
\end{equation}
When we numerically evaluate the expectation values, we find that $a^4t_2/t$ is slightly different from $b$. 
This is probably owing to the difference in the evaluation of integrals involving $V_0({\bm r}-{\bm R})$, 
and we think that this difference is not important. 
Therefore, $\Delta \chi$ qualitatively corresponds to $\chi_2$. 
To confirm this correspondence, we estimate the second term in (\ref{MatsuuraT2}) as follows:
\begin{equation}
\begin{split}
&\sim \frac{1}{2a^4} \biggl[ s\ \langle (R_x y- R_y x)^2 \rangle_{{\rm p}\pi {\rm p}\pi}^{(0)}
\langle V_0({\bm r}-{\bm R}) \rangle_{{\rm p}\pi {\rm p}\pi}^{(0)} \cr
&\qquad \qquad -\langle (R_x y- R_y x)^2 \rangle_{{\rm p}\pi {\rm p}\pi}^{(0)}
\langle V_0({\bm r}-{\bm R})\rangle_{R, {\rm p}\pi {\rm p}\pi}^{(0)} \biggr]. \cr
&\sim \frac{1}{2a^4} \langle (R_x y- R_y x)^2 \rangle_{{\rm p}\pi {\rm p}\pi}^{(0)}\times t,
\label{EstimateT2}
\end{split}
\end{equation}
where we have used the definition of $t$ in (\ref{tdef1s}).
Using the expectation value in (\ref{x2in2ppi}) and $R_x^2+R_y^2=a^2$, (\ref{EstimateT2}) becomes 
\begin{equation}
\frac{1}{2a^4} t\left( R_x^2 \langle y^2\rangle_{{\rm p}\pi {\rm p}\pi}^{(0)} 
+  R_y^2 \langle x^2\rangle_{{\rm p}\pi {\rm p}\pi}^{(0)} \right) = \frac{3 a_{\rm B}^{*2}t}{a^2}. 
\end{equation}
In this estimation, $t_2$ becomes 
\begin{equation}
t_2 \sim \frac{\hbar^2}{2ma^4} 
\left( \langle x^2 + y^2 \rangle_{R, {\rm p}\pi {\rm p}\pi}^{(0)} 
- s \langle x^2 + y^2 \rangle_{{\rm p}\pi {\rm p}\pi}^{(0)} \right) + \frac{3 a_{\rm B}^{*2}t}{a^2}. 
\end{equation}
Substituting this expression into (\ref{Matsuura}), we recover the results 
in (\ref{Chi2Final}) and (\ref{Chi2FinalB}).

Finally, let us discuss the relationship of the results of this work 
to the recent work by Gao {\it et al.}\cite{Gao}
[Note that their formula in the time-reversal symmetric case is almost equivalent to the present 
formula in (\ref{FinalChi}) except for the numerical factor of $\chi_{\rm FS}$.\cite{OgaFuku1}]
In Ref.~17, the orbital susceptibility for gapped graphene was calculated. 
It was claimed that the contribution of energy polarization (which corresponds to our $\chi_{\rm FS}/2$) 
and that of the Van Vleck susceptibility (which corresponds to $\chi_{\rm inter}$) 
vanish in the case of gapped graphene. 
Furthermore, Langevin-type and geometrical susceptibilities (which correspond to $\chi_{\rm occ}$) 
give a symmetric contribution with respect to the sign change of $\mu$. 
These results are different from the present case of (gapless) graphene. 
Thus, it is an interesting future problem to examine the model of gapped graphene, 
in which the potential $V({\bm r})$ is noncentrosymmetric. 
In such cases, orbital magnetization\cite{Thon} and the connection to the so-called \lq\lq Berry" curvature 
can be discussed.

In summary, we calculated the orbital susceptibility for graphene or for an electron system on a  
two-dimensional honeycomb lattice based on a newly developed general formula. 
Our result contains all the contributions up to the first order with respect to the overlap integrals 
between the nearest-neighbor atomic orbitals. 
In contrast to the previous studies, we found the additional contributions of $\chi_{\rm occ:1}$ and $\chi_2$. 
In particular, the former gives asymmetry with respect to the sign change of the chemical potential. 
Furthermore, the physical origin of the new contribution, $\chi_2$, is clarified in terms of the 
corrections to the Peierls phase argument. 
Furthermore, 
because of the nature of the two-band model, $\chi_{\rm inter}$ gives a finite contribution to $\chi$, 
which is different from the case of the single-band model discussed in the preceding paper.\cite{OgataII}
The relative phase $\theta_{\bm k}$ between the atomic orbitals on A and B sublattices plays important roles. 
Interestingly, $\theta_{\bm k}$ is directly related to the chirality of the two Dirac cones 
and also to the integral called the interband Berry connection. 

\bigskip\noindent
{\bf Acknowledgments}

We thank H.\ Fukuyama, F.\ Pi\'echon, H.\ Matsuura, 
I.\ Proskurin, T.\ Kariyado, Y.\ Fuseya, T.\ Mizoguchi and N.\ Okuma 
for very fruitful discussions. 
This work was supported by a Grant-in-Aid for Scientific Research on 
\lq\lq Multiferroics in Dirac electron materials'' (No.\ 15H02108).

\appendix

\section{Overlap integrals and matrix elements}

Various kinds of expectation values defined in eqs.\ (\ref{defOverlap}), (\ref{tdef1s}), 
and (\ref{ExpValueZero}) can be obtained analytically for $\phi_{{\rm p}\pi} ({\bm r})$. 
Without loss of generality, we assume ${\bm R}=(a,0,0)$. 
Then using a change of coordinates, 
$\xi= r+r_b, \eta = r-r_b$ with $r=|{\bm r}|, r_b = |{\bm r}-{\bm R}|$,\cite{Mulliken} 
we obtain 
\begin{equation}
\begin{split}
\langle r_\perp^2 \rangle_{R, {\rm p}\pi {\rm p}\pi}^{(0)}  &= 
6a_{\rm B}^{*2} \left( 1+{\tilde p}+\frac{13{\tilde p}^2}{21} + \frac{2{\tilde p}^3}{7} + 
\frac{3{\tilde p}^4}{35} + \frac{4{\tilde p}^5}{315} \right)\ e^{-{\tilde p}}, \cr 
\langle r_\parallel^2 \rangle_{R, {\rm p}\pi {\rm p}\pi}^{(0)}  &= 
6a_{\rm B}^{*2} \left( 1+{\tilde p}+\frac{3{\tilde p}^2}{7} + \frac{2{\tilde p}^3}{21} + 
\frac{{\tilde p}^4}{105} \right)\ e^{-{\tilde p}},
\label{IntResultPpi}
\end{split}
\end{equation}
for the p$_\pi$ orbital with ${\tilde p}=a/2a_{\rm B}^*$, 
where $r_\perp (r_\parallel)$ denotes the coordinate in the 
direction perpendicular (parallel) to $\bm R$. 
In a similar way, we obtain $s$, $t_0$ and $c_{{\rm p}\pi}$ in Eq.~(\ref{OverlapIntPpi}). 
When we set $\tilde p=0$, we have 
\begin{equation}
\langle x^2 \rangle^{(0)}_{{\rm p}\pi {\rm p}\pi}=\langle y^2 \rangle^{(0)}_{{\rm p}\pi {\rm p}\pi}=6a_{\rm B}^{*2}. 
\label{x2in2ppi}
\end{equation}

Expectation values in terms of $\Phi_{{\rm p}\pi}({\bm r})$ defined in Eq.~(\ref{Odefinition}) 
can be easily obtained from 
$\langle {\cal O} \rangle_{{\rm p}\pi {\rm p}\pi}^{(0)}$. 
For example,\cite{OgataII}
\begin{equation}
\langle {\cal O} \rangle_{{\rm p}\pi {\rm p}\pi} = \langle {\cal O} \rangle_{{\rm p}\pi {\rm p}\pi}^{(0)} + O(s^2).
\end{equation}
and 
\begin{equation}
\begin{split}
\langle {\cal O}({\bm r}) \rangle_{R, {\rm p}\pi {\rm p}\pi} 
= &\langle {\cal O}({\bm r}) \rangle_{R, {\rm p}\pi {\rm p}\pi} ^{(0)} 
-\frac{s}{2} \langle {\cal O}({\bm r}) \rangle_{{\rm p}\pi {\rm p}\pi} ^{(0)} \cr
&-\frac{s}{2} \langle {\cal O}({\bm r}+{\bm R}) \rangle_{{\rm p}\pi {\rm p}\pi} ^{(0)} + O(s^2). 
\end{split}
\end{equation}
Using these relations, we have 
\begin{equation}
\langle x^2 +y^2 \rangle_{R, {\rm p}\pi {\rm p}\pi} = \langle x^2+y^2 \rangle_{R, {\rm p}\pi {\rm p}\pi}^{(0)} 
- s \langle x^2+y^2 \rangle_{{\rm p}\pi {\rm p}\pi}^{(0)} -\frac{s}{2} a^2, 
\label{ExpR2inR}
\end{equation}
and $\langle x^2+y^2 \rangle_{R, {\rm p}\pi {\rm p}\pi}^{(0)}$ is obtained from 
$\langle r_\parallel^2 \rangle_{R, {\rm p}\pi {\rm p}\pi}^{(0)}+
\langle r_\perp^2 \rangle_{R, {\rm p}\pi {\rm p}\pi}^{(0)}$.

Next we calculate the matrix elements that appear in $\chi_{\rm inter}$, $\chi_{\rm FS}$,  
and $\chi_{\rm occ}$. 
For example, using the $\bm k$-derivatives of $u_{{\rm p}\pi {\bm k}}^{\pm}({\bm r})$ 
in (\ref{Ukderiv}), we obtain the interband matrix element as
\begin{equation}
\begin{split}
&\int u_{{\rm p}\pi {\bm k}}^{\pm \dagger}({\bm r})
\frac{\partial u_{{\rm p}\pi {\bm k}}^{\mp}({\bm r})}{\partial k_x} d{\bm r} 
= \pm \frac{\theta_x}{2} \cr
&\qquad -\frac{1}{2} \sum_{{\bm R}} e^{-i\theta_{\bm k}} e^{-i{\bm k}\cdot {\bm R}}
\langle x+ \frac{\theta_x}{2} \rangle_{R,{\rm p}\pi {\rm p}\pi} \cr
&\qquad +\frac{1}{2} \sum_{{\bm R}} e^{i\theta_{\bm k}} e^{i{\bm k}\cdot {\bm R}}
\langle x - R_x - \frac{\theta_x}{2} \rangle^*_{R,{\rm p}\pi {\rm p}\pi} 
 + O(s^2), 
\label{BerryInt}
\end{split}
\end{equation}
where ${\bm R}={\bm R}_{{\rm A}i} -{\bm R}_{{\rm B}j}$ and 
we have used abbreviations such as $\theta_x=\partial \theta_{\bm k}/\partial k_x$. 
The terms with the ${\bm R}$-summation in (\ref{BerryInt}) originate from the integrals 
between the nearest-neighbor sites, which are in the first order of the overlap integrals. 
As in the 1s orbital case, we can show that 
$\langle 1 \rangle_{R,{\rm p}\pi {\rm p}\pi}
= \langle x \rangle_{R,{\rm p}\pi {\rm p}\pi} 
= \langle y \rangle_{R,{\rm p}\pi {\rm p}\pi}
=0$.\cite{OgataII}
As a result, we obtain Eq.~(\ref{Berry}). 
In the same way, we can show that ($\mu, \nu=x, y$) 
\begin{equation}
\int u_{{\rm p}\pi {\bm k}}^{\pm\dagger} \frac{\partial u_{{\rm p}\pi {\bm k}}^{\pm}}{\partial k_\mu} d{\bm r}=0, 
\label{IntFormula10}
\end{equation}
\begin{equation}
\begin{split}
&\int \frac{\partial u_{{\rm p}\pi {\bm k}}^{\pm\dagger}}{\partial k_\mu} 
\frac{\partial u_{{\rm p}\pi {\bm k}}^{\pm}}{\partial k_\nu} d{\bm r} 
=\langle x_\mu x_\nu \rangle_{{\rm p}\pi{\rm p}\pi}
+\frac{1}{4}\theta_\mu \theta_\nu \cr
&\qquad \mp {\rm Re} \sum_{{\bm R}} e^{-i\theta_{\bm k}} e^{-i{\bm k}\cdot {\bm R}}
\langle x_\mu x_\nu \rangle_{R,{\rm p}\pi {\rm p}\pi} + O(s^2),
\label{IntFormula11}
\end{split}
\end{equation}
\begin{equation}
\int \frac{\partial u_{{\rm p}\pi {\bm k}}^{\pm \dagger}}{\partial k_\mu} 
\frac{\partial u_{{\rm p}\pi {\bm k}}^{\mp}}{\partial k_\nu} d{\bm r} =O(s), 
\label{IntFormula12}
\end{equation}
\begin{equation}
\int u_{{\rm p}\pi {\bm k}}^{\pm \dagger}
\frac{\partial^2 u_{{\rm p}\pi {\bm k}}^{\mp}}{\partial k_\mu \partial k_\nu} d{\bm r} 
= \pm \frac{1}{2} \theta_{\mu\nu} + O(s). 
\label{IntFormula13}
\end{equation}

Furthermore, using the general relation 
\begin{equation}
\frac{\partial H_{\bm k}}{\partial k_\mu} \Uellp + H_{\bm k}  \frac{\partial \Uellp}{\partial k_\mu}
= \frac{\partial \Eellp}{\partial k_\mu} \Uellp + \Eellp \frac{\partial \Uellp}{\partial k_\mu},
\label{IntFormula14}
\end{equation}
which is derived from the ${\bm k}$ derivative of the equation for $\Uellp$ in (\ref{UellEq}), 
we obtain\cite{OgaFuku1}
\begin{equation}
\int \Uell^\dagger \frac{\partial H_{\bm k}}{\partial k_\mu} \Uellp d{\bm r} 
= \frac{\partial \Eellp}{\partial k_\mu} \delta_{\ell \ell'} + 
(\Eellp - \Eell) \int \Uell^\dagger \frac{\partial \Uellp}{\partial k_\mu} d{\bm r}. 
\label{IntFormula15}
\end{equation}
From this relation, we have
\begin{equation}
\int u_{{\rm p}\pi {\bm k}}^{\pm \dagger} \frac{\partial H_{\bm k}}{\partial k_\mu} 
u_{{\rm p}\pi {\bm k}}^{\pm} d{\bm r} 
= \frac{\partial \varepsilon_{{\rm p}\pi\bm k}^\pm}{\partial k_\mu} = \pm \varepsilon_\mu,
\label{IntFormula16}
\end{equation}
and 
\begin{equation}
\int u_{{\rm p}\pi {\bm k}}^{\pm\dagger} \frac{\partial H_{\bm k}}{\partial k_\mu} 
u_{{\rm p}\pi {\bm k}}^{\mp} d{\bm r} 
=\mp 2\varepsilon_{\bm k} \int u_{{\rm p}\pi {\bm k}}^{\pm\dagger} 
\frac{\partial u_{{\rm p}\pi {\bm k}}^{\mp}}{\partial k_\mu} d{\bm r} 
= -\varepsilon_{\bm k} \theta_\mu,
\label{IntFormula17}
\end{equation}
where we have used the relation (\ref{Berry}). 

Furthermore, from the $k_\mu$ and $k_\nu$ derivatives of Eq.~(\ref{UellEq}), we obtain
 \begin{equation}
 \begin{split}
 &\left( \frac{\hbar^2}{m} \delta_{\mu\nu} -  \frac{\partial^2 \Eell}{\partial k_\mu \partial k_\nu} 
 \right) \delta_{\ell \ell'} 
 + \int \Uellp^\dagger \left( 
 \frac{\partial H_{\bm k}}{\partial k_\mu} - \frac{\partial \Eell}{\partial k_\mu} \right)
 \frac{\partial \Uell}{\partial k_\nu} d{\bm r} \cr 
 &+ \int \Uellp^\dagger \left( 
 \frac{\partial H_{\bm k}}{\partial k_\nu} - \frac{\partial \Eell}{\partial k_\nu} \right) 
 \frac{\partial \Uell}{\partial k_\mu} d{\bm r} 
 + \left( \Eellp - \Eell \right) \int \Uellp^\dagger \frac{\partial^2 \Uell}{\partial k_\mu \partial k_\nu}
 d{\bm r} = 0.
 \label{AppB3p}
 \end{split}
 \end{equation}
From this relation, we can show that\cite{OgaFuku1}
\begin{equation}
\begin{split}
\int \frac{\partial u_{{\rm p}\pi {\bm k}}^{\pm\dagger}}{\partial k_\mu} 
\frac{\partial H_{\bm k}}{\partial k_\nu} u_{{\rm p}\pi {\bm k}}^{\pm} d{\bm r}
&=\int u_{{\rm p}\pi {\bm k}}^{\pm \dagger}  \frac{\partial H_{\bm k}}{\partial k_\nu} 
 \frac{\partial u_{{\rm p}\pi {\bm k}}^{\pm}}{\partial k_\mu} d{\bm r} \cr
&= \frac{1}{2} \left( \pm \varepsilon_{\mu\nu} -\frac{\hbar^2}{m} \delta_{\mu\nu}\right),
\label{IntFormula18}
\end{split}
\end{equation}
\begin{equation}
\int u_{{\rm p}\pi {\bm k}}^{\mp \dagger}  \frac{\partial H_{\bm k}}{\partial k_x} 
\frac{\partial u_{{\rm p}\pi {\bm k}}^{\pm}}{\partial k_x} d{\bm r} 
= - \frac{1}{2}\left( \varepsilon_{\bm k} \theta_{xx} + \varepsilon_x \theta_x \right) + O(s^2),
\label{IntFormula19}
\end{equation}
and
\begin{equation}
\begin{split}
&\int u_{{\rm p}\pi {\bm k}}^{\mp \dagger}  \frac{\partial H_{\bm k}}{\partial k_x} 
\frac{\partial u_{{\rm p}\pi {\bm k}}^{\pm}}{\partial k_y} d{\bm r} +  
\int u_{{\rm p}\pi {\bm k}}^{\mp \dagger}  \frac{\partial H_{\bm k}}{\partial k_x} 
\frac{\partial u_{{\rm p}\pi {\bm k}}^{\pm}}{\partial k_y} d{\bm r} \cr
&= - \varepsilon_{\bm k} \theta_{xy} - \frac{1}{2} 
\left( \varepsilon_x \theta_y + \varepsilon_y \theta_x \right) + O(s^2),
\label{IntFormula20}
\end{split}
\end{equation}
where we have used the relations in (\ref{Berry}) and (\ref{IntFormula13}).

\section{Density of states for graphene}

In the case of graphene or a honeycomb lattice, the Brillouin zone is also 
a honeycomb with a size of $8\sqrt{3}\pi^2/9a^2$ 
with $a$ being the nearest-neighbor distance, and the system area is $L^2=3\sqrt{3}a^2 N/2$ 
with $N$ being the total number of unit cells. 
Therefore, the density of states per area is obtained as
\begin{equation}
D(\mu) = \frac{2\sqrt{3}}{9Na^2} \sum_{{\bm k},\pm} \delta(\pm \varepsilon_{\bm k}-\mu) 
= \frac{1}{4\pi^2} \iint_{\rm B.Z.} dk_x dk_y 
\delta ( \pm t |\gamma_{\bm k}| - \mu ).
\label{TriangleIntDOS}
\end{equation}
After performing the $k_y$-integral and the change of variable $x=\cos^2 \sqrt{3}k_xa/2$, we obtain
\begin{equation}
\begin{split}
D(\mu) &=\frac{4|\mu|}{3\sqrt{3}\pi^2 t^2a^2} \int_{0}^{1} 
\frac{\theta(16x - (4x+1-\mu^2/t^2)^2)dx}{\sqrt{x}\sqrt{1-x}\sqrt{16x - (4x+1-\mu^2/t^2)^2}}\cr
&=\frac{|\mu|}{3\sqrt{3}\pi^2 t^2a^2} \int_{0}^{1} 
\frac{\theta( (\alpha-x)(x-\beta) )}{\sqrt{x(1-x)(\alpha-x)(x-\beta)}} dx,
\end{split}
\end{equation}
with
\begin{equation}
\alpha = \left( \frac{1+|\mu|/t}{2} \right)^2, \qquad 
\beta = \left( \frac{1-|\mu|/t}{2} \right)^2. 
\end{equation}
Finally, using the formula\cite{OgataII,Gradzyen}
\begin{equation}
\int_{c}^{b} \frac{dx}{\sqrt{(a-x)(b-x)(x-c)(x-d)}} = \frac{2}{\sqrt{(a-c)(b-d)}} K(q),
\label{Formula10}
\end{equation}
with $q = \sqrt{{(a-d)(b-c)}/{(a-c)(b-d)}}$
for $a>b>c>d$, we obtain Eq.~(\ref{DOSgra}).

At the bottom (or top) of the band ($\mu = \pm 3t$), $\kappa$ becomes zero and  
using $K(0)=\pi/2$, we obtain 
\begin{equation}
D(\pm 3t) = \frac{1}{3\pi ta^2}.
\end{equation}
This is consistent with the fact that the model is equivalent to the free electrons with 
an effective mass $m^*=2\hbar^2/3ta^2$ at the band edge. 

$D(\mu)$ has a diverging peak at $\mu=\pm 2t$, which corresponds to the 
van Hove singularity. From the analytical form of (\ref{DOSgra}), we obtain
\begin{equation}
D(\mu\sim 1) = \frac{2}{3\sqrt{3} \pi^2 ta^2} \ln \frac{16}{|\mu-1|^3}, 
\end{equation}
where we have used the fact that $\kappa$ behaves as 
\begin{equation}
\kappa^2 \sim 1-|\mu-1|^3/4,
\end{equation}
near $\mu=1$ and $K(\kappa) \sim \ln (4/\sqrt{1-\kappa^2})$ as $\kappa\rightarrow 1$. 
On the other hand, when $\mu\sim 0$, we can show that 
\begin{equation}
D(\mu\sim 0) = \frac{4|\mu|}{9 \pi t^2a^2}. 
\end{equation}
This is consistent with the density of states of the two massless Dirac electrons 
around the two Dirac points with velocity $3ta/2$ [see Eq.~(\ref{DiracLimit})].

\section{Calculation of $\chi_{\rm inter}$ and $\chi_{\rm FS}$ for graphene}

First, we calculate the integrals in the second line of (\ref{ChiInterPpi}) for $\chi_{\rm inter}$. 
Using the $\bm k$-derivatives of $u_{{\rm p}\pi {\bm k}}^{\pm}$ 
in (\ref{Ukderiv}) and the relation $\hat L_z \phi_{{\rm p}_\pi}({\bm r})=0$, we can show that
\begin{equation}
\begin{split}
&\frac{\partial H_{\bm k}}{\partial k_y}  \frac{\partial u_{{\rm p}\pi {\bm k}}^{\pm}}{\partial k_x} 
=\frac{\partial H_{\bm k}}{\partial k_x}  \frac{\partial u_{{\rm p}\pi {\bm k}}^{\pm}}{\partial k_y}
-\left( \frac{\theta_x}{2} \frac{\partial H_{\bm k}}{\partial k_y} 
-  \frac{\theta_y}{2} \frac{\partial H_{\bm k}}{\partial k_x} \right) \cr
&\times \frac{C^{\pm}}{\sqrt{2}}\left\{ 
e^{ \frac{i}{2}\theta_{\bm k}} \varphi_{{\rm A}{\bm k}}^{{\rm ortho}} ({\bm r}) \pm 
e^{-\frac{i}{2}\theta_{\bm k}} \varphi_{{\rm B}{\bm k}}^{{\rm ortho}} ({\bm r}) \right\} \cr
&+\frac{s}{2} \sum_{\bm R} \left( R_x \frac{\partial H_{\bm k}}{\partial k_y} 
- R_y \frac{\partial H_{\bm k}}{\partial k_x} \right) \cr
&\times  \frac{C^{\pm}}{\sqrt{2}} \left\{ e^{i{\bm k}{\bm R}} e^{\frac{i}{2}\theta_{\bm k}} 
\varphi_{{\rm B}{\bm k}}^{{\rm ortho}} ({\bm r}) \pm 
e^{-i{\bm k}{\bm R}} e^{-\frac{i}{2}\theta_{\bm k}} 
\varphi_{{\rm A}{\bm k}}^{{\rm ortho}} ({\bm r}) \right\} +O(s^2), 
\label{ChiInterMat1}
\end{split}
\end{equation}
with ${\bm R}={\bm R}_{{\rm A}i} -{\bm R}_{{\rm B}j}$. 
Note that the terms with the $\bm R$-summation appear since 
$\Phi_{{\rm p}\pi}({\bm r}-{\bm R}_{{\rm A}i})$ contains the 
nearest-neighbor orbitals, $-(s/2)\phi_{{\rm p}\pi}({\bm r}-{\bm R}_{{\rm B}j})$ 
in addition to $\phi_{{\rm p}\pi}({\bm r}-{\bm R}_{{\rm A}i})$ [see Eq.\ (\ref{OrthoNormal})]. 
The $\bm R$-summation in (\ref{ChiInterMat1}) can be carried out as 
\begin{equation}
\sum_{{\bm R}\ne 0} R_x {\rm e}^{-i{\bm k}{\bm R}} 
= i \frac{\partial \gamma_{\bm k}}{\partial k_x} 
= i \frac{\partial}{\partial k_x} \left( |\gamma_{\bm k}| e^{i\theta_{\bm k}} \right) 
= \left( \frac{i\varepsilon_x}{t} -  |\gamma_{\bm k}| \theta_x\right) e^{i\theta_{\bm k}}.
\label{RsumFormula1}
\end{equation}
Also, using the definition of $u_{{\rm p}\pi {\bm k}}^{\pm}$ in (\ref{UellGraphene}), we 
can rewrite (\ref{ChiInterMat1}) as
\begin{equation}
\begin{split}
\frac{\partial H_{\bm k}}{\partial k_y}  \frac{\partial u_{{\rm p}\pi {\bm k}}^{\pm}}{\partial k_x} 
=&\frac{\partial H_{\bm k}}{\partial k_x}  \frac{\partial u_{{\rm p}\pi {\bm k}}^{\pm}}{\partial k_y}
\mp \left( \frac{\theta_x}{2} \frac{\partial H_{\bm k}}{\partial k_y} 
-  \frac{\theta_y}{2} \frac{\partial H_{\bm k}}{\partial k_x} \right)
\left( 1\pm s|\gamma_{\bm k}| \right) u_{{\rm p}\pi {\bm k}}^{\mp} \cr
&\pm \left( \frac{s\varepsilon_x}{2t} \frac{\partial H_{\bm k}}{\partial k_y}  
- \frac{s\varepsilon_y}{2t} \frac{\partial H_{\bm k}}{\partial k_x}  \right) 
u_{{\rm p}\pi {\bm k}}^{\pm} +O(s^2).
\label{ChiInterMat1b}
\end{split}
\end{equation}
Note that the second term on the right-hand side has $u_{{\rm p}\pi {\bm k}}^{\mp}$, 
which means that the bonding and antibonding orbitals are exchanged in this term. 
Using this relation, we can see that the matrix element of $\chi_{\rm inter}$ 
in (\ref{ChiInterPpi}) becomes
\begin{equation}
\begin{split}
& \int \left[ \mp \frac{\theta_x}{2} 
\left( 1\pm s|\gamma_{\bm k}| \right) u_{{\rm p}\pi {\bm k}}^{\mp\dagger}
\pm \frac{s \varepsilon_x}{2t} u_{{\rm p}\pi {\bm k}}^{\pm\dagger} \right]  
\frac{\partial H_{\bm k}}{\partial k_y}  \Uellp d{\bm r} \cr
&\pm \int \frac{\partial u_{{\rm p}\pi {\bm k}}^{\pm\dagger}}{\partial k_x} 
\varepsilon_y
\Uellp d{\bm r} 
+O(s^2) -(x\leftrightarrow y).
\label{ChiInterMat2}
\end{split}
\end{equation}

Let us first calculate the case with $\Uellp=u_{{\rm p}\pi {\bm k}}^{\mp}$.
In this case, with the help of (\ref{IntFormula16}) and (\ref{IntFormula17}), 
the first term in (\ref{ChiInterMat2}) becomes
$\theta_x \varepsilon_y/2+O(s^2)$, 
which cancels with the second term using the relation (\ref{Berry}). 
This cancellation means that the numerator in $\chi_{\rm inter}$ in (\ref{ChiInterPpi}) 
when $\Uellp=u_{{\rm p}\pi {\bm k}}^{\mp}$ is 
proportional to the fourth order of overlap integrals, i.e., $O(s^4)$. 

Next, we consider the case with $\varepsilon_{\ell'} \ne \varepsilon_{{\rm p}\pi\bm k}^\mp$. 
In this case, with the help of the general relation in (\ref{IntFormula14}),  (\ref{ChiInterMat2}) becomes
\begin{equation}
\begin{split}
&-\int \biggl[ \mp \frac{\theta_x}{2} 
\left( 1\pm s|\gamma_{\bm k}| \right) (\varepsilon_{{\rm p}\pi}^\mp({\bm k}) - \Eellp)
u_{{\rm p}\pi {\bm k}}^{\mp\dagger} \cr
&\pm \frac{s \varepsilon_x}{2t} (\varepsilon_{{\rm p}\pi}^\pm({\bm k}) - \Eellp) 
u_{{\rm p}\pi {\bm k}}^{\pm\dagger} \biggr] 
\frac{\partial \Uellp}{\partial k_y} d{\bm r}  
\pm \varepsilon_y \int \frac{\partial u_{{\rm p}\pi {\bm k}}^{\pm\dagger}}{\partial k_x} 
\Uellp d{\bm r}  -(x\leftrightarrow y) \cr 
&= (\varepsilon_{{\rm p}\pi}^\pm({\bm k}) - \Eellp) \int \biggl[ \mp \frac{\theta_x}{2} 
\left( 1\pm s|\gamma_{\bm k}| \right) \frac{\partial u_{{\rm p}\pi {\bm k}}^{\mp\dagger}}{\partial k_y} 
\pm \frac{s \varepsilon_x}{2t} \frac{\partial u_{{\rm p}\pi {\bm k}}^{\pm\dagger}}{\partial k_y} \biggr] 
\Uellp d{\bm r} \cr 
&+ \varepsilon_{\bm k} \theta_x
\left( 1\pm s|\gamma_{\bm k}| \right) \int \frac{\partial u_{{\rm p}\pi {\bm k}}^{\mp\dagger}}{\partial k_y} 
\Uellp d{\bm r} 
\pm \varepsilon_y \int \frac{\partial u_{{\rm p}\pi {\bm k}}^{\pm\dagger}}{\partial k_x} 
\Uellp d{\bm r}- (x\leftrightarrow y),
\label{ChiInterMat3}
\end{split}
\end{equation}
where $O(s^2)$ terms are neglected, and we have used
$H_{\bm k} u_{{\rm p}\pi {\bm k}}^{\pm} = \varepsilon_{{\rm p}\pi}^\pm({\bm k}) u_{{\rm p}\pi {\bm k}}^{\pm}$, 
$\varepsilon_{{\rm p}\pi}^\pm({\bm k}) -\varepsilon_{{\rm p}\pi}^\mp({\bm k}) = \pm 2\varepsilon_{\bm k}$,
and the relations
\begin{equation}
\begin{split}
&\int u_{{\rm p}\pi {\bm k}}^{\pm\dagger} \Uellp d{\bm r} = 0, \cr
&\int u_{{\rm p}\pi {\bm k}}^{\pm\dagger} \frac{\partial \Uellp}{\partial k_y} d{\bm r}
=- \int \frac{\partial u_{{\rm p}\pi {\bm k}}^{\pm\dagger}}{\partial k_y} \Uellp d{\bm r}. 
\end{split}
\end{equation}
The latter relation is obtained by partial derivative of the former (orthogonality condition).\cite{OgaFuku1} 

The factor $ (\varepsilon_{{\rm p}\pi\bm k}^\pm - \Eellp)$ in the first term in (\ref{ChiInterMat3})
cancels with the denominator of $\chi_{\rm inter}$ in (\ref{ChiInterPpi}). 
As a result, the $\ell'$-summation in $\chi_{\rm inter}$ can be carried out using the 
completeness condition for $\Uellp$. 
(A similar method was used in I\cite{OgaFuku1} when obtaining 
the $f$-sum rule.) 
Taking account of the fact that the second term in (\ref{ChiInterMat3}) is in the first order 
with respect to the overlap integrals, we obtain
\begin{equation}
\begin{split}
&\chi_{\rm inter} = -\frac{e^2}{\hbar^2 c^2} \sum_{\pm} f(\varepsilon_{\bm k}^\pm )
\biggl[ \frac{\theta_x^2}{4} (1\pm s|\gamma_{\bm k}|)^2  
\int \frac{\partial u_{{\rm p}\pi {\bm k}}^{\mp \dagger}}{\partial k_y} 
\frac{\partial H_{\bm k}}{\partial k_y} u_{{\rm p}\pi {\bm k}}^{\mp} d{\bm r} \cr
&\qquad \mp \frac{1}{2} \varepsilon_{\bm k} \theta_x^2 
\int \frac{\partial u_{{\rm p}\pi {\bm k}}^{\mp \dagger}}{\partial k_y} 
\frac{\partial u_{{\rm p}\pi {\bm k}}^{\mp}}{\partial k_y}  d{\bm r} + (3\ {\rm terms}) \biggr] +O(s^2),
\end{split}
\end{equation}
where (3 terms) represents the terms in which subscripts $(xxyy)$ are changed to 
$(xyyx), (yxxy)$, and $(yyxx)$ with minus signs for $(xyyx)$ and $(yxxy)$. 
Here, we have used (\ref{Berry}), (\ref{IntFormula10}), (\ref{IntFormula14}), (\ref{IntFormula12}), 
(\ref{IntFormula17}), (\ref{IntFormula19}), and (\ref{IntFormula20}). 
Finally, using (\ref{IntFormula11}) and the complex conjugate of (\ref{IntFormula18}), 
we obtain (\ref{GrChiInterFin}). 

Next we calculate $\chi_{\rm FS}$. In a similar way to obtain (\ref{ChiInterMat2}), we can show that
\begin{equation}
\begin{split}
&\int \frac{\partial u_{{\rm p}\pi {\bm k}}^{\pm\dagger}}{\partial k_y} 
\frac{\partial H_{\bm k}}{\partial k_x}  
\frac{\partial u_{{\rm p}\pi {\bm k}}^{\pm}}{\partial k_y} d{\bm r} 
-\int \frac{\partial u_{{\rm p}\pi {\bm k}}^{\pm\dagger}}{\partial k_x} 
\frac{\partial H_{\bm k}}{\partial k_y}  
\frac{\partial u_{{\rm p}\pi {\bm k}}^{\pm}}{\partial k_y} d{\bm r} \cr
&=\int \left[ \pm \frac{\theta_x}{2} 
\left( 1\pm s|\gamma_{\bm k}| \right) u_{{\rm p}\pi {\bm k}}^{\mp\dagger}
\mp \frac{s \varepsilon_x}{2t} u_{{\rm p}\pi {\bm k}}^{\pm\dagger} \right]  
\frac{\partial H_{\bm k}}{\partial k_y}  \frac{\partial u_{{\rm p}\pi {\bm k}}^{\pm}}{\partial k_y} 
d{\bm r} \cr
&-\int \left[ \pm \frac{\theta_y}{2} 
\left( 1\pm s|\gamma_{\bm k}| \right) u_{{\rm p}\pi {\bm k}}^{\mp\dagger}
\mp \frac{s \varepsilon_y}{2t} u_{{\rm p}\pi {\bm k}}^{\pm\dagger} \right]  
\frac{\partial H_{\bm k}}{\partial k_x}  \frac{\partial u_{{\rm p}\pi {\bm k}}^{\pm}}{\partial k_y} 
d{\bm r}. 
\label{ChiFSMat1}
\end{split}
\end{equation}
With the help of (\ref{IntFormula18})-(\ref{IntFormula20}), (\ref{ChiFSMat1}) becomes
\begin{equation}
\mp \frac{\varepsilon_{\bm k}}{4} \left( \theta_x \theta_{yy} - \theta_y \theta_{xy} \right) 
\pm \frac{\hbar^2 s}{4mt} \varepsilon_x + O(s^2).
\end{equation}
Here, we have used the transformation in (\ref{ChiInterMat1b}) to use the 
relation in (\ref{IntFormula20}) for the last term. 
Finally, collecting all the terms in $\chi_{\rm FS}$, we obtain (\ref{GrChiFSFin}). 

\section{Proof of Eq.~(\ref{XRelation})}

Generally we have
\begin{equation}
\sum_{{\bm R}\ne 0} R_x^2 {\rm e}^{-i{\bm k}{\bm R}} 
= - \frac{\partial^2 \gamma_{\bm k}}{\partial k_x^2} 
= - \frac{\varepsilon_{xx} + 2i \varepsilon_x \theta_x + i\varepsilon_{\bm k} \theta_{xx}
- \varepsilon_{\bm k} \theta_x^2}{t} e^{i\theta_{\bm k}}. 
\end{equation}
However, $\sum_{{\bm R}\ne 0} (R_x^2 + R_y^2) {\rm e}^{-i{\bm k}{\bm R}}=a^2 \gamma_{\bm k}
=a^2 \varepsilon_{\bm k}e^{i\theta_{\bm k}}/t$ holds. 
Therefore, taking the summation of $R_x^2 + R_y^2$ and taking its real part, we obtain
\begin{equation}
\varepsilon_{\bm k} \theta_x^2 + \varepsilon_{\bm k} \theta_x^2 - \varepsilon_{xx} - \varepsilon_{yy} 
= a^2 \varepsilon_{\bm k}.
\end{equation}
On the other hand, its imaginary part gives
\begin{equation}
2 \varepsilon_x \theta_x + 2 \varepsilon_y \theta_y 
+ \varepsilon_{\bm k} \theta_{xx} + \varepsilon_{\bm k} \theta_{yy} = 0. 
\end{equation}

\section{Comparison between results of G\'omez-Santos {\it et al.} and Raoux {\it et al.}}

The orbital susceptibility obtained by Raoux {\it et al.} is given by [see Eq.~(30) in Ref.\cite{Piechon}.]
\begin{equation}
\begin{split}
\chi^{({\rm R})} &= \frac{e^2}{6\hbar^2 c^2} \sum_{{\bm k}, \pm} 
\biggl[ (U_1-V_1-4V_2) \left( f'(\pm \varepsilon_{\bm k}) 
\mp \frac{f(\pm \varepsilon_{\bm k})}{\varepsilon_{\bm k}} \right) \cr
&\qquad\qquad\quad \pm U_2 \frac{f(\pm \varepsilon_{\bm k})}{\varepsilon_{\bm k}} 
\mp V_1 \varepsilon_{\bm k} f''(\pm \varepsilon_{\bm k}) \biggr], 
\label{PieChi}
\end{split}
\end{equation}
with 
\begin{equation}
\begin{split}
U_1 &= \frac{1}{\varepsilon_{\bm k}^2} \left\{ ({\bm f}_{xx} \cdot {\bm f}) ({\bm f}_{yy} \cdot {\bm f})
- ({\bm f}_{xy} \cdot {\bm f})^2 \right\},  \cr
U_2 &= {\bm f}_{xx} \cdot {\bm f}_{yy} - {\bm f}_{xy} \cdot {\bm f}_{xy}, \cr
V_1 &= \frac{1}{\varepsilon_{\bm k}^2} (\varepsilon_y {\bm f}_{x} - \varepsilon_x {\bm f}_{y})^2, \qquad
V_2 = \frac{1}{\varepsilon_{\bm k}^4} (({\bm f}_{x} \times {\bm f}_{y})\cdot {\bm f})^2,
\end{split}
\end{equation}
where the subscripts means the partial derivatives with respect to $\bm k$. 
Here, ${\bm f}$ is defined as ${\bm f}=(t {\rm Re}\ \gamma_{\bm k}, t {\rm Im}\ \gamma_{\bm k}, 0)$
in the present notation, with Re (Im) representing the real (imaginary) part. 
Therefore, $|{\bm f}| = t|\gamma_{\bm k}|$. 

In the present notation, $\theta_{\bm k}$ is represented as
\begin{equation}
\theta_{\bm k} = {\rm tan}^{-1} \left( {{\rm Im}\ \gamma_{\bm k}}/{{\rm Re}\ \gamma_{\bm k}} \right).
\end{equation}
Using this relation, we can rewrite $U_1, U_2, V_1$, and $V_2$ as follows:
\begin{equation}
\begin{split}
U_1 &= 
\left( \varepsilon_{xx} - \varepsilon_{\bm k} \theta_x^2 \right) 
\left( \varepsilon_{yy} - \varepsilon_{\bm k} \theta_y^2 \right) 
-\left( \varepsilon_{xy} - \varepsilon_{\bm k} \theta_x \theta_y \right)^2, \cr
V_1 &= (\theta_x \varepsilon_{y} - \theta_y \varepsilon_{x})^2, \qquad
V_2 =0, \cr
U_1- &V_1-U_2 = -\varepsilon_{\bm k}^2 (\theta_{xx} \theta_{yy} - \theta_{xy}^2) \cr
&-2\varepsilon_{\bm k} \left( \varepsilon_{x} \theta_x \theta_{yy} 
+ \varepsilon_{y} \theta_y \theta_{xx} - \varepsilon_{x} \theta_y \theta_{xy} 
- \varepsilon_{y} \theta_x \theta_{xy} \right).
\end{split}
\end{equation}
Finally substituting them into (\ref{PieChi}) and using integration by parts, we obtain 
an orbital susceptibility equal to G\`omez-Santos and Stauber's result in (\ref{Gomez}),
or equivalently $\chi_{\rm LP}+\chi_1$.

\def\journal#1#2#3#4{#1 {\bf #2}, #3 (#4)}
\def\PR{Phys.\ Rev.}
\def\PRB{Phys.\ Rev.\ B}
\def\PRL{Phys.\ Rev.\ Lett.}
\def\JPSJ{J.\ Phys.\ Soc.\ Jpn.}
\def\PTP{Prog.\ Theor.\ Phys.}
\def\JPCS{J.\ Phys.\ Chem.\ Solids}
\def\RMP{Rev.\ Mod.\ Phys.}

\end{document}